\documentclass[%
reprint,twocolumn,longbibliography,superscriptaddress]{revtex4-2}
\usepackage{graphicx}
\usepackage{dcolumn}

\usepackage{bm}
\usepackage[utf8]{inputenc}
\usepackage{epsfig}
\usepackage{float}
\usepackage{dcolumn}   
\usepackage{array}
\usepackage{amssymb}   
\usepackage{amsmath}
\usepackage{amstext}
\usepackage{mathtools}
\usepackage{threeparttable}
\usepackage{siunitx} 
\usepackage{lipsum}  
\usepackage{mathrsfs}
\usepackage{physics}
\usepackage{xcolor}
\usepackage[draft]{todonotes}
\usepackage{float}
\usepackage[normalem]{ulem}

\usepackage [colorlinks=true,allcolors=blue]{hyperref}

\def\@element#1#2\@nil{%
  #1%
 \if\relax#2\relax\else\MakeLowercase{#2}\fi}
\makeatother

\begin{document}
\preprint{APS/123-QED}
\title{Incommensurate Magnetic Order in Hole-Doped Infinite-layer Nickelate Superconductors}

\author{Yajun Zhang}
\email{zhangyajun@lzu.edu.cn} %
\affiliation{Key Laboratory of Mechanics on Disaster and Environment in Western China Attached to The Ministry of Education of China, Lanzhou University, Lanzhou 730000 Gansu, China}%
\affiliation{Department of Mechanics and Engineering Science, College of Civil Engineering and Mechanics, Lanzhou University, Lanzhou 730000 Gansu, China}%
\author{ Xu He}
\affiliation{Institute of Condensed Matter and Nanosciences, Universit\'e Catholique de Louvain, 1348 Louvain-la-Neuve, Belgium}%
\author{Jingtong Zhang}
\affiliation{Theoretical Materials Physics, Q-MAT, CESAM, Universit\'e de Li\`ege, B-4000 Li\`ege, Belgium}%
\affiliation{Department of Engineering Mechanics and Key Laboratory of Soft Machines and Smart Devices of Zhejiang Province, Zhejiang University, 38 Zheda Road, Hangzhou 310027, China}%

\author{Jie Wang }
\affiliation{Department of Engineering Mechanics and Key Laboratory of Soft Machines and Smart Devices of Zhejiang Province, Zhejiang University, 38 Zheda Road, Hangzhou 310027, China}%

\author{ Philippe Ghosez}
\affiliation{Theoretical Materials Physics, Q-MAT, CESAM, Universit\'e de Li\`ege, B-4000 Li\`ege, Belgium}%

\begin{abstract}
Magnetism and superconductivity are closely entangled, elucidating the
magnetic interactions in nickelate superconductors is at the heart of
understanding the pairing mechanism. Our first-principles and spin-wave theory
calculations highlight that NdNiO$_2$ is in the vicinity of a transition between
a quasi-two-dimensional (2D) antiferromagnetic (AFM) state and a three-dimensional (3D) C-AFM state. Both states could accurately reproduce the experimentally measured magnetic
excitation spectra, which was previously explained in terms
of a 2D model. We further reveal that hole doping stabilizes an incommensurate
(IC) spin state and the IC wave vector increases continuously. Direct links between hole doping, magnetization, exchange constants, and magnetic order are established, revealing that the competition between first-neighbor and third-neighbor in-plane magnetic interactions is the key for the IC magnetic order.

\end{abstract}

\maketitle



Infinite-layer nickelates $R$NiO$_2$ ($R$ = rare-earth ion) have recently attracted a lot of attention owing to the discovery of superconductivity \cite{li2019superconductivity,osada2020superconducting,hepting2020electronic,lu2021magnetic,wang2021isotropic,leonov2020lifshitz,ryee2020induced,kapeghian2020electronic,botana2020similarities,liu2020electronic,zhang2021magnetic,been2021electronic,fowlie2022intrinsic,jiang2020critical,wan2021exchange,li2020superconducting,goodge2021doping,chen2022magnetism,krishna2020effects,nomura2019formation,xia2022dynamical,bernardini2022geometric,tam2022charge,krieger2022charge,rossi2022broken,chen2022charge}. This new class of hole-doped superconductors not only fueled theoretical and experimental research seeking for new superconductors, but also acts as a unique and valuable system for revealing the key ingredients responsible for the pairing mechanism of superconductivity. Magnetic interactions are commonly believed to play essential roles for the emergence of superconductivity \cite{moriya2000spin,kastner1998magnetic,j2006magnetic}, as a result, rationalizing the existence and origin of magnetism is critical for the understanding of superconductivity mechanism. 

\par
\setlength{\parskip}{0em} 
For undoped system, it is known that cuprates are characterized by
strongly two-dimensional (2D) antiferromagnetic (AFM) interactions
\cite{endoh1988static} while ferropnictides exhibit essentially
three-dimensional (3D) magnetism  \cite{zhao2009spin}. Recent resonant inelastic
X-ray scattering (RIXS) experiment suggests a 2D magnetic coupling in NdNiO$_2$
similar to that of CaCuO$_2$ \cite{lu2021magnetic}. The demonstration of cuprate-like 2D
magnetic dimensionality is on the base of the unnoticeable dispersion along the
specific (0.25, 0, 0.25)-(0.25, 0, 0.39) path, while there is no information
about the magnetic spectrum along other out-of-plane paths. At the opposite, 3D
G-AFM or C-AFM order with notable interlayer magnetic coupling has been
theoretically predicted as the magnetic ground state \cite{liu2020electronic,leonov2020lifshitz,ryee2020induced,zhang2021magnetic,kapeghian2020electronic,botana2020similarities}. This questions the exact dimensionality and origin of the magnetic interactions in undoped infinite-layer nickelates.

\par
\setlength{\parskip}{0em} 
For hole-doped superconductors, incommensurate (IC) spin order is one of the
most intriguing and generic features of cuprates and iron superconductors,
and play an important role in the superconducting phase diagram
\cite{moriya2000spin,kastner1998magnetic,j2006magnetic}. Interestingly, the
superconducting phase diagram of infinite-layer nickelates is strikingly similar
to that of cuprates \cite{li2020superconducting}. Given the similar electronic
configuration and magnetic excitations in their parent compounds, it is natural
to wonder if there is similar IC magnetic order in hole-doped nickelate
superconductors. A complete understanding of the magnetic ground state in
undoped and hole-doped infinite-layer nickelates will timely provide the support
for unravelling the superconducting mechanism.

\par
\setlength{\parskip}{0em} 
In this letter, we report extensive first-principles studies of the magnetic
ground state in undoped and hole-doped NdNiO$_2$ and its
link to electronic degrees of freedom. By comparison with the RIXS
experiments, we reveal that the spin-wave dispersions derived from a quasi-2D AFM
state and a weakly 3D C-AFM state agree both very well with the experimental
data. However, the 3D AFM model exhibits additional out-of-plane dispersions along some wave
vector directions not probed experimentally. Moreover, we demonstrate that hole
doping has profound impacts on the magnetization, exchange interactions, and
magnetic order, thereby yielding a commensurate-IC transition, reminiscent of
the spin fluctuations in cuprate and iron superconductors. The amplitude of
incommensurability increases monotonically with increasing doping level x from the lightly doped side to the overdoped regime. We highlight
that the main roles of hole doping are to control the sign and ratio of the in-plane first-neighbor and third-neighbor magnetic interactions, which eventually determine the magnetic ground state and the incommensurability.

\begin{figure} 
	\includegraphics[width=0.8\columnwidth]{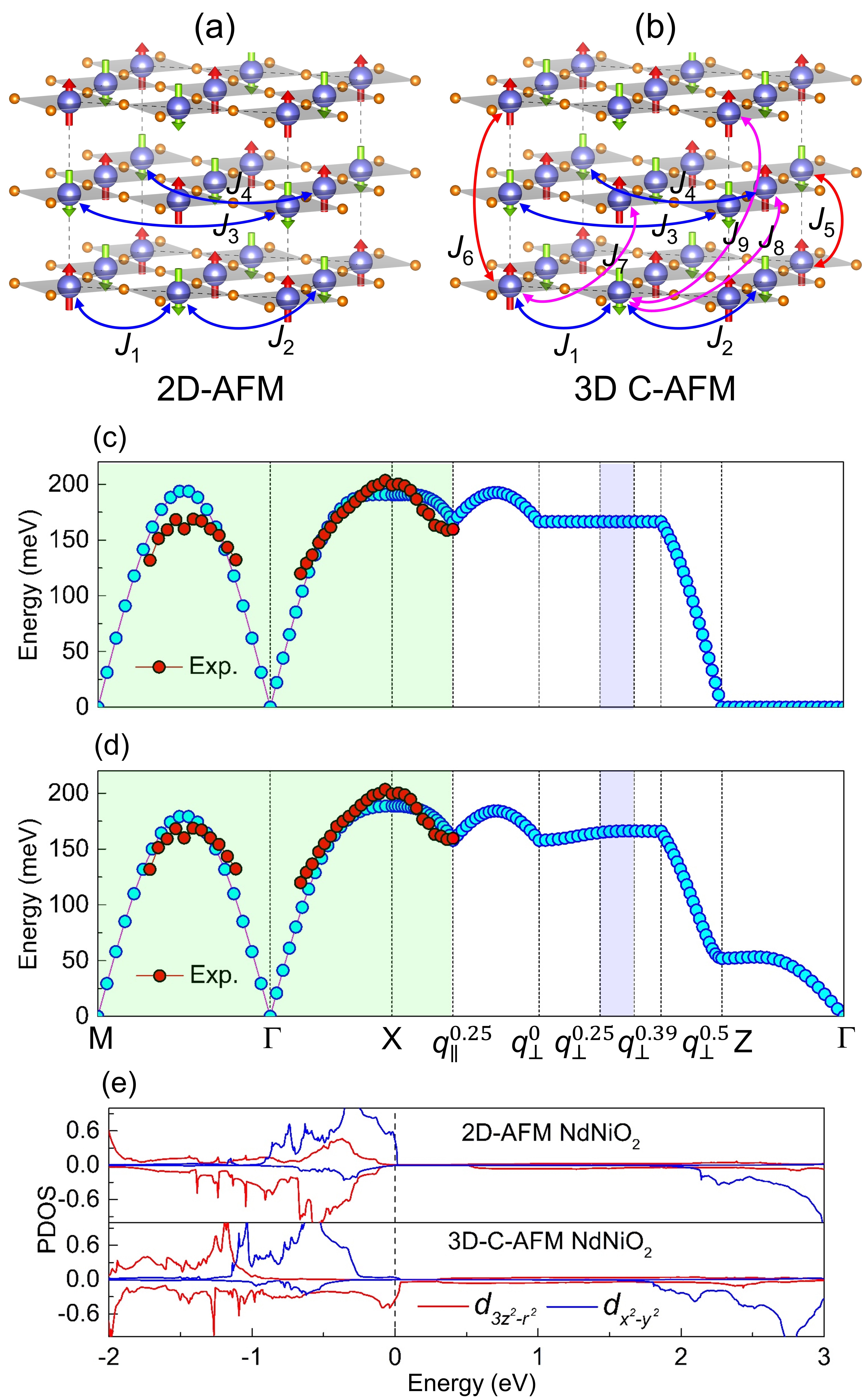}
    \caption{Magnetic dimensionality of NdNiO$_2$ and the electronic origin. Schematic representation of (a) 2D-AFM and (b) 3D C-AFM states with different neighboring exchange interactions. The simulated adiabatic spin-wave dispersions based on (c) 2D-AFM model with Perdew-Burke-Ernzerhof (PBE) \cite{perdew1996generalized} + U (U = 0.83 eV) and (d) 3D C-AFM model with PBE + U (U = 0.84 eV) along different path of the Brillouin zone of the tetragonal phase. The experimental results \cite{lu2021magnetic} in red circle are also shown for comparison. Here, M = (0.5, 0.5, 0), $\Gamma =$ (0, 0, 0), X = (0.5, 0, 0), $q^{0.25}_\rVert$ = (0.5, 0.25, 0), $q^{0}_\perp$ = (0.25, 0, 0), $q^{0.25}_\perp$ = (0.25, 0, 0.25), $q^{0.39}_\perp$ = (0.25, 0, 0.39), $q^{0.5}_\perp$ = (0.25, 0, 0.5), and Z = (0, 0, 0.5). A segment of the path $q^{0.25}_{\perp}-q^{0.39}_{\perp}$ with background in blue is taken in Ref.~\cite{lu2021magnetic} to show the flatness of the dispersion along the out-of-plane direction. (e) Projected density of states (PDOS) of Ni atom for 2D-AFM and 3D C-AFM NdNiO$_2$. The Fermi level denoted by the dash line is set to zero energy.}
    \label{fig:Fig4:JT_couplings}
\end{figure}

We perform first-principles calculations using density functional theory (DFT) plus U \cite{dudarev1998electron} for the ground state properties (Figs. S1 and S2 in
the Supplemental Material) \cite {{note1}}, TB2J code \cite{he2021tb2j} to extract orbital-dependent exchange constant and calculate the spin-wave vector, and SpinW code \cite{ toth2015linear} to calculate the spin-wave spectra. The technical details, discussion of ground state and magnetic dimensionality, and selection of Hubbard U are provided in the Supplemental Material \cite {{note1}}. \nocite {kresse1993ab,blochl1994projector,sun2015strongly,fu2018applicability,monkhorst1976special,togo2008first,marzari2012maximally,mostofi2014updated,korotin2015calculation,zhang2020self}

\par
\setlength{\parskip}{0em}

Our central finding in undoped NdNiO$_2$ is displayed in Figs. 1(a)-1(d), which shows the comparison between the experimental measurements and the predicted magnetic excitation spectrum obtained from a 2D-AFM state [Fig. 1(a)] and from a 3D C-AFM state with non-negligible out-of-plane ferromagnetic (FM) coupling [Fig. 1(b)]. Obviously, the magnetic dispersions obtained from both our 2D (Fig. 1(c)) and 3D (Fig. 1(d)) models with a relative small Hubbard U accurately reproduce the RIXS results \cite{lu2021magnetic}, including the absence of noticeable dispersion along the (0.25, 0, 25)-(0.25, 0, 0.39) path. However, our 3D model highlights pronounced dispersion along the $\Gamma$-Z direction not captured by our 2D model. This result clearly demonstrates that the restricted path measured experimentally does not allow to distinguish unambiguously between the quasi-2D AFM and weakly 3D magnetic state.

\par
\setlength{\parskip}{0em} 
Interestingly, the two possible states are very close in energy as shown in Fig. S2(a) \cite {{note1}} while they have strikingly different electronic structure. In the quasi-2D AFM state, the Fermi level is dominated by the Ni $d_{x^2-y^2}$ electrons (Fig. 1(e)). In the weakly 3D C-AFM state, the itinerant Ni $d_{3z^2-r^2}$ orbital intersects the Fermi level, while the $d_{x^2-y^2}$ electron is strongly localized below the Fermi level (Fig. 1(e)). The orbital contributions of the exchange constants are listed in Table SII \cite {{note1}}. We see clearly that the first-neighbor in-plane magnetic coupling ($J_1$) in both the quasi-2D AFM and weakly 3D C-AFM states mainly originates from the contribution of the localized $d_{x^2-y^2}$ orbital. Nevertheless, the presence of first-neighbor out-of-plane FM coupling ($J_5$) in the weakly 3D C-AFM state emerges from the interactions between itinerant Ni $d_{3z^2-r^2}$ electrons. Consequently, a direct relationship between Fermi surface and magnetic order is established: i) the localized $d_{x^2-y^2}$ electrons alone are responsible for the 2D magnetic dimensionality; ii) the itinerant $d_{3z^2-r^2}$ and localized $d_{x^2-y^2}$ electrons account together for the 3D magnetic state in C-AFM NdNiO$_2$. As both magnetic states show good consistency with current experiments, the determination of magnetic dimensionality calls for further experimental measurements along other out-of-plane directions.

\begin{figure}
	\centering\includegraphics[width=0.95\columnwidth]{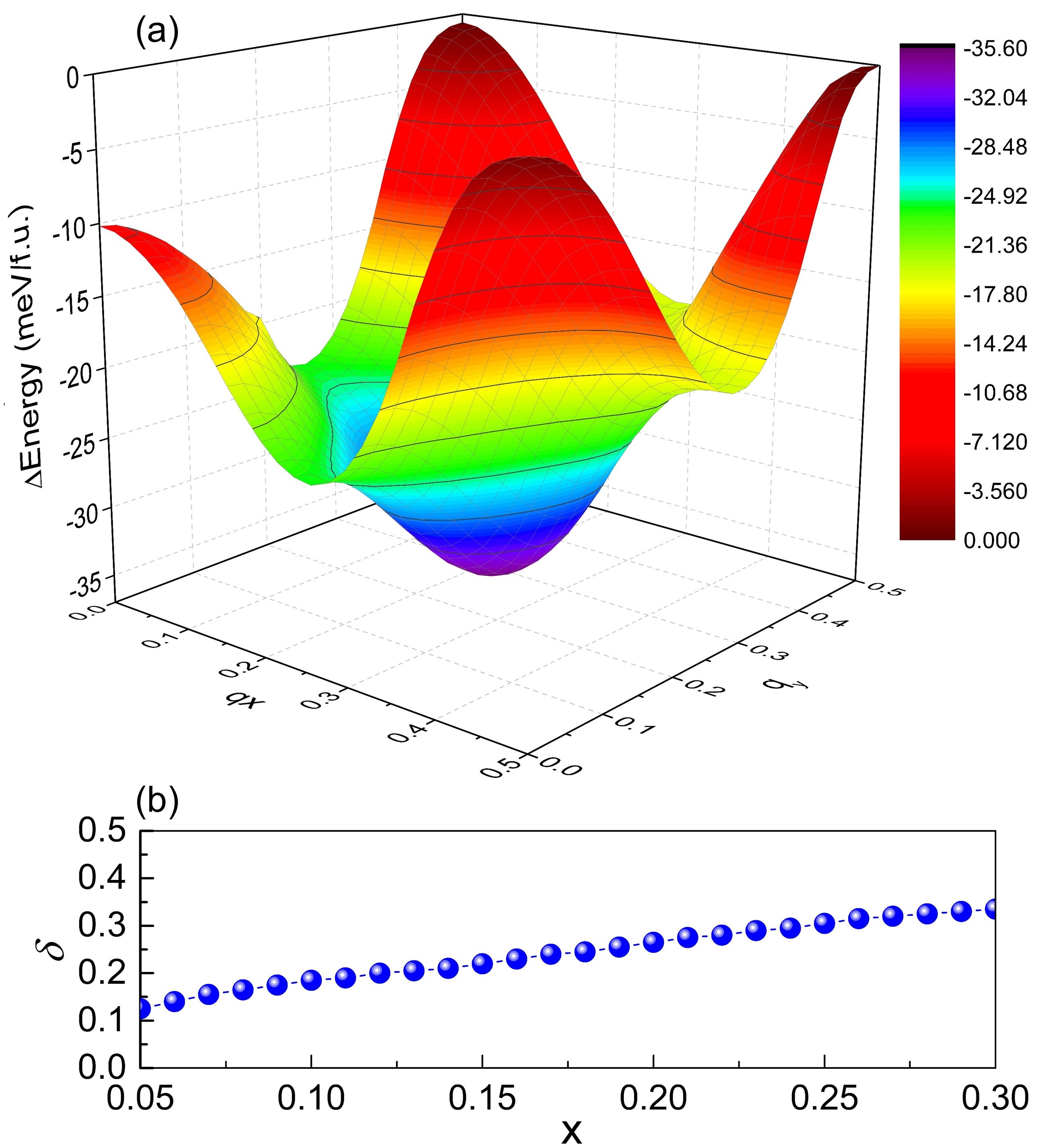}
    \caption{Verification of IC magnetic ground state in hole-doped NdNiO$_2$.
      (a) The energy difference $\Delta$E between different spin spiral states
      and 2D-AFM state for Sr$_{0.18}$Nd$_{0.82}$NiO$_2$. (b) Doping
      concentration dependent incommensurability $\delta$.}
    \label{fig:delta}
\end{figure}
\par
\setlength{\parskip}{0em} 
Having obtained suitable description of the magnetic interactions in the undoped NdNiO$_2$, we now explore further the possible occurrence of an IC spin state in hole-doped $P4/mmm$ NdNiO$_2$ by comparing the total energies E($q$) of the spin spiral states with different wave vector $q$ using the generalized Bloch theorem (GBT) \cite{sandratskii1998noncollinear}. Figure 2(a) presents the energy difference $\Delta$E between the spin spiral states and 2D-AFM state at 0.18 hole doping, in the middle of the superconducting regime (0.125$\leq$x$\leq$0.25) \cite{li2020superconducting}. It is worth noting that the initial spin state (2D-AFM or 3D C-AFM) in the parent NdNiO$_2$ does not affect the present discussions on the IC magnetic order (see Fig. S3) \cite {{note1}}.

\par 
\setlength{\parskip}{0em} 
It is evident from Fig. 2(a) that the state with q$_x$=$q_y$=0.26 is the most
energetically favorable phase, which is a definitive signature of IC magnetic
order. Using the same method, the incommensurability $\delta$ ($\delta$ = 0.5-q)
from lightly doped to overdoped states is summarized in Fig. 2(b).
Interestingly, $\delta$ varies dramatically with Sr doping and two features are
obviously analogous to that of La$_{2-x}$Sr$_x$CuO$_4$ \cite{j2006magnetic}: i)
the IC state occurs before the transition to the superconducting phase; ii) the
incommensurability $\delta$ increases almost linearly with doping when it starts to
across the superconducting phase boundary. Overall, the similarity of
the IC magnetic state in NdNiO$_2$ and cuprate superconductors over a wide range
of doping indicates that the IC spin fluctuations are robust and a general feature
of hole-doped superconductors, which could be naturally extended to infinite-layer nickelates.

\par
\setlength{\parskip}{0em} 
While the IC spin order in nickelate superconductors is directly confirmed by the GBT, this approach alone is inadequate to identify the microscopic origin of magnetic incommensurability. Here, an alternative and easier method based on the calculated exchange constants with magnetic force theorem (MFT) \cite{liechtenstein1987local} from the FM state is employed to determine the incommensurability $\rm {\delta}$. This method is only exact when the spin state is close to the DFT reference state but it remains a good approximation when the electronic state is insensitive to the spin order. Therefore, we use this method to analyze the range of doping 0.16$\leq$x$\leq$0.3, where the $J_1$ is close to the FM alignment. In Fig. 3(a), we display $\delta$ in terms of the doping concentration. Clearly, the monotonic increase and $\delta$ values agree reasonably well with the results from GBT, attesting that the MFT gives a good approximation. The consistent results from two entirely different approaches not only provide convincing evidence for the IC nature of magnetic order, but also demonstrate that our predictions are not the artifact of specific method.

\par
\setlength{\parskip}{0em} 
To reveal the microscopic origin of IC spin state, we first compare the change of
exchange constants as a function of doping. From Fig. S4 \cite {{note1}}, we notice that the
in-plane magnetic interactions are mostly dominated by the first-neighbor,
second-neighbor, third-neighbor, and fourth neighbor exchange interactions,
which are schematically shown in Fig. 1(a). Figure 3(b) displays the variation of five exchange constants as a function of hole doping. It is obvious that $J_1$ evolves much faster than the other interactions and the strength shows almost linear decrease (increase) in the AFM (FM) region. Moreover, the first-neighbor magnetic coupling undergoes an AFM-FM transition at the hole concentration of x=0.15, which is accompanied by the FM-AFM transition of the second-neighbor and third-neighbor magnetic couplings. Strikingly, the strength of $J_3$ becomes comparable to that of $J_1$ at relative high doping concentration.
\par
\setlength{\parskip}{0em}

Next, we elucidate how the IC spin state and $\delta$ connect to the variation of these magnetic interactions. With a deeper insight into the individual impact of each magnetic interaction, we find that $\delta$ depends crucially on the first-neighbor and third-neighbor magnetic interactions while is only slightly affected by the second-neighbor and fourth-neighbor magnetic interactions (see Fig. S5) \cite {{note1}}. In details, we notice that $J_3$, when it is AFM, competes with the in-plane commensurate FM and AFM states which favors a FM alignment of the in-plane third-neighbor magnetic moments. The balance of the competition is therefore mainly between the $J_1$ and $J_3$, and is decisive for the incommensurability. Indeed, the IC state is suppressed when the AFM $J_3$ decreases, and eventually becomes commensurate when $J_3$ turns FM.

\par
\setlength{\parskip}{0em}

\par
\setlength{\parskip}{0em}


To further clarify the role of the first-neighbor and third-neighbor magnetic
interactions in the origin of the IC state, we analyze a simplified $J_1-J_3$ Heisenberg model by eliminating the other exchange interactions: 
\begin{equation}
E=\sum_{R\in \text{1N}}J_1S(0)S(R) + \sum_{R\in \text{3N}}J_3S(0)S(R)
\end{equation}
where $R\in \text{1N}$ and $R\in \text{3N}$ mean the first-neighbor and third-neighbor, respectively. For a spin spiral state, wave vector \(\vec{q}\) can be written as \(S_q(R)=Se^{i\vec{q}\cdot\vec{R}}\), we can evaluate the energy as a function of \(\vec{q}\) and find its minimum.

\begin{figure}
\centering\includegraphics[width=0.95\columnwidth]{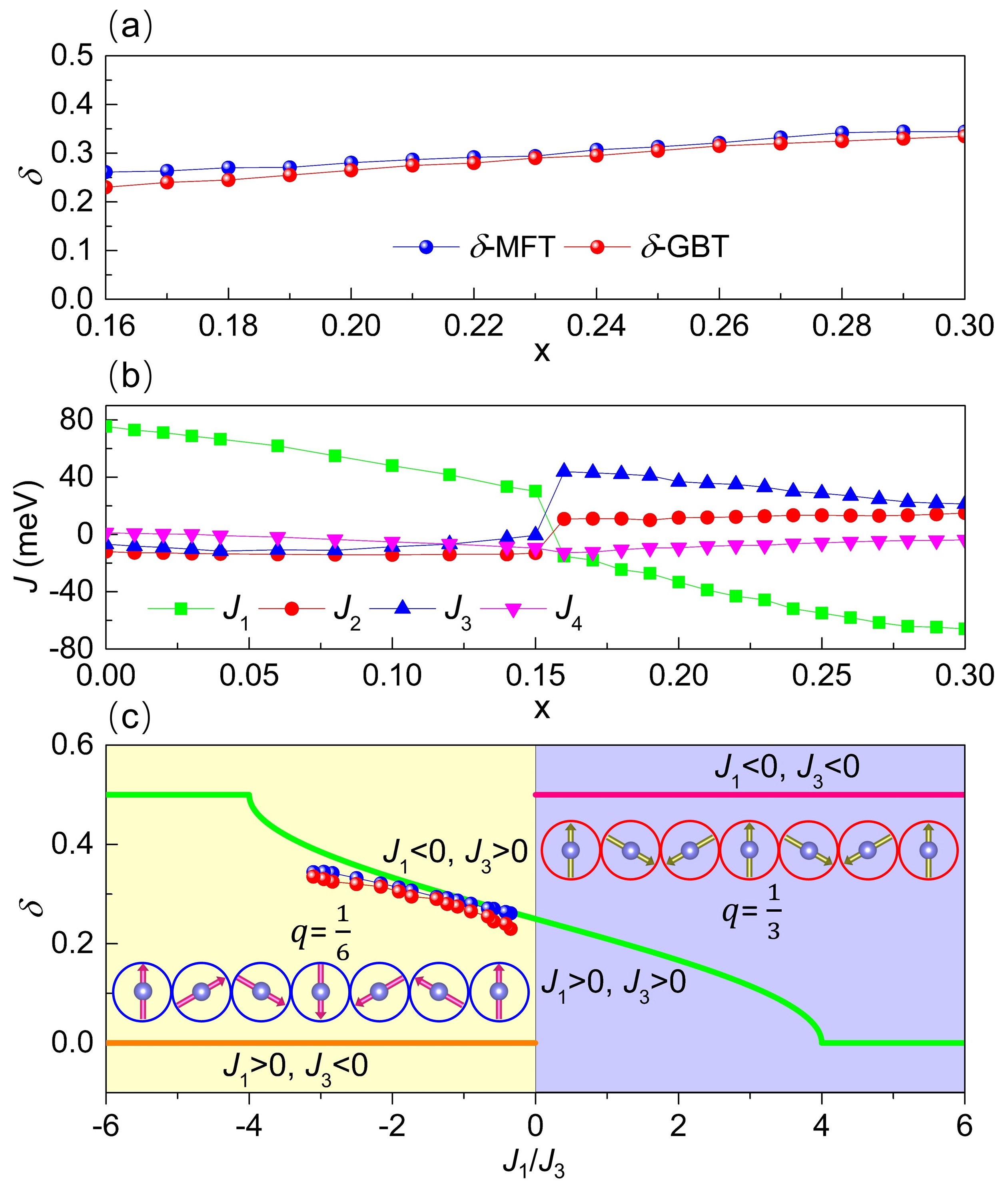}
\caption{The microscopic origin of the IC magnetic state. (a) the comparison of
  incommensurability $\delta$ from the MFT and GBT. (b) The variation of
  dominant exchange constants as a function of doping level $x$. The $J$ values
    are computed from the lowest-energy collinear-spin DFT calculations, specifically, the G-AFM (x$\leq$0.15) and the
    A-AFM states (x$\geq$0.16). Positive values correspond to AFM coupling. (c) The
    analytical solution of $\delta$ from the $J_1-J_3$ model. The results from (a) are also plot together for comparison with the $J_1/J_3$ ratio from MFT. The sketches of spin waves ($q$=1/6 and 1/3) highlight the region of FM (resp. AFM) in-plane first-neighbor magnetic interaction. }
    \label{fig:qvsJ}
\end{figure}

With a FM $J_3$ ($J_3 <0$), the magnetic structure is commensurate and the
order depends on $J_1$. The wave vector is $q = 0$ when $J_1$ is FM, and is $q=\frac{1}{2}$ when $J_1$ is AFM. With an AFM $J_3$ ($J_3>0$), the wave vector $q$ depends on the
ratio of $\frac{J_1}{J_3}$.

\begin{equation}
  \label{eq:qvsJ}
  q=
  \begin{dcases}
    0,  & \text{if\quad}   -\cfrac{J_1}{4J_3}>1 \\
    \frac{1}{2},  & \text{if\quad}  -\cfrac{J_1}{4J_3}<-1 \\
 \frac{1}{2\pi}   \arccos(\cfrac{-J_1}{4J_3}) ,  &  \text{if\quad} -1<-\cfrac{J_1}{4J_3}<1 \\
  \end{dcases}
\end{equation}
here, $q = q_x = q_y$, and $q$ has the unit of $2\pi/a$. The dependence of $\delta$
on the $J_1/J_3$ is shown in Fig.~\ref{fig:qvsJ} (c). The results obtained
from both the GBT and MFT (with all the exchange constants) methods agree reasonably well with the numerical solutions, indicating the effectiveness of the $J_1-J_3$ model.

\par
\setlength{\parskip}{0em} 
\par
\setlength{\parskip}{0em} 
\begin{figure}
\centering\includegraphics[width=0.9\columnwidth]{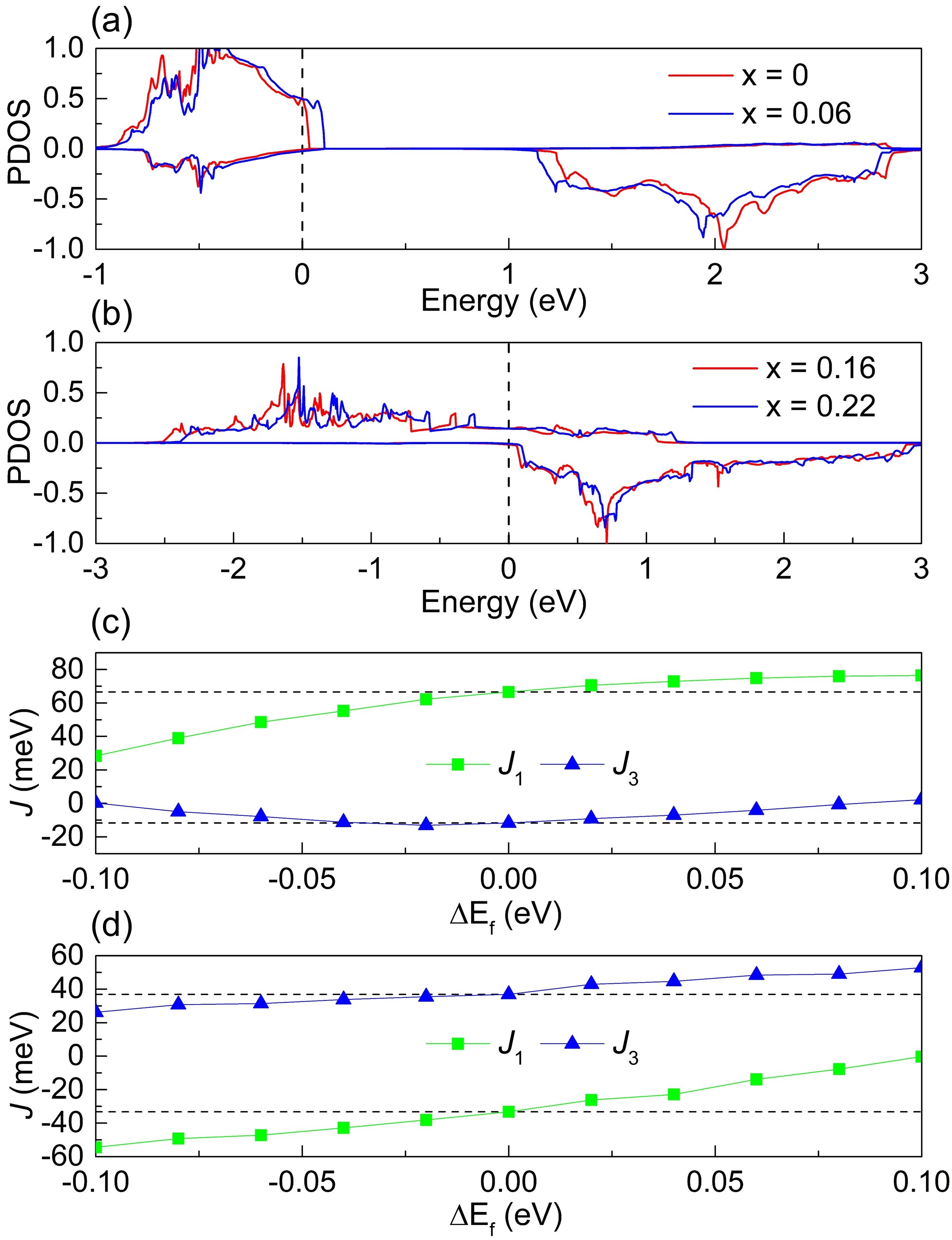}
    \caption{The hole doping-magnetization-magnetic interactions relationship.
      The PDOS for $d_{x^2-y^2}$ band of Ni atom in Sr$_{x}$Nd$_{1-x}$NiO$_2$ with (a) G-AFM state for
      undoped and lightly doped states and (b) A-AFM state for moderately doped
      states. The effect of the shift of Fermi energy (E$_f$) on the
      $J_1$ and $J_3$ for (c)
      Sr$_{0.04}$Nd$_{0.96}$NiO$_2$ and (d) Sr$_{0.2}$Nd$_{0.8}$NiO$_2$.
      In (c) and (d), the dashed lines are the values without Fermi energy shifting.}
    \label{fig:electron-}
\end{figure}

Nowadays, it is proposed that the IC spiral state coexists and couples
with superconductivity. However, the microscopic origin of this specific phase
remains controversial. Fermi surface topology \cite{si1993comparison}, dipolar
distortion \cite{shraiman1989spiral}, next-nearest neighbor hopping
\cite{sushkov2004superconducting}, stripe-domain \cite{tranquada1995evidence},
and interaction between holes \cite{chubukov1995magnetic} have been proposed
to be responsible for the IC spiral state. Here, we propose
a different perspective by establishing a direct relation between the electronic
structure, the exchange constant, and the IC state. Then, the key question is
how this mechanism is actually connected with the $d_{x^2-y^2}$ band which is
the origin of the pairing interaction \cite{scalapino1995case}.

To answer this question and gain further insight into the relationship between the magnetic
interaction and the electronic states, the exchange constants from different electronic states are compared. Figures 4(a) and 4(b) display the hole-doping dependence of PDOS of Ni $d_{x^2-y^2}$ band for doped NdNiO$_2$. Consistent with recent experiments \cite{osada2020superconducting}, we found that Sr doping has strong effect on the Ni $d$ orbital. From the small to large doping concentrations, one can observe clearly that the spin polarization gradually decreases and the Fermi level shifts down continuously in line with previous works \cite{chen2022magnetism,krishna2020effects}. In order to elucidate how the magnetic interactions are influenced by the reduction of spin polarization, we artificially shift the Fermi level of Sr$_{0.04}$Nd$_{0.96}$NiO$_2$ (Fig. 4(c)) and Sr$_{0.2}$Nd$_{0.8}$NiO$_2$ (Fig. 4(d)) and calculate the exchange constants at each state. Strikingly, both $J_1$ and $J_3$ are dominated by the magnetic interactions from the $d_{x^2-y^2}$ bands, as evident in Table SIII \cite {{note1}}. As the Fermi level shifts down, for Sr$_{0.04}$Nd$_{0.96}$NiO$_2$ with first-neighbor AFM interactions, $J_1$ decreases more quickly than the increase of FM $J_3$. According to the phase diagram in Fig. 3(c), the decrease of $J_1$/$J_3$ ratio directly triggers the commensurate-IC transition and causes the increase of incommensurability $\delta$. In terms of Sr$_{0.2}$Nd$_{0.8}$NiO$_2$ with first-neighbor FM interactions, $J_1$ increases much faster than the weakening of AFM $J_3$ as the Fermi level shifts down. This results in the increase of $J_1$/$J_3$ ratio and consequently increases the incommensurability $\delta$ on the basis of the phase diagram shown in Fig. 3(c). These scenarios unambiguously confirm that the weakened magnetization of the $d_{x^2-y^2}$ band induced by hole doping suppresses the first-neighbor AFM interaction, and enhances the first-neighbor FM interactions with sufficient doping, which is responsible for the occurrence of IC spin state and continuous increase of $\delta$ upon increasing doping. Our results thus establish a clear connection between the hole doping, magnetization of $d_{x^2-y^2}$ band, variation of exchange constant, and the IC spin state. 
\par
\setlength{\parskip}{0em} 
It should be noticed that in our calculations with GBT and MFT, a homogenous spin/charge density on the Ni sites is assumed. In high-T$_c$ superconductors, other states like spin/charge density wave \cite{emery1993frustrated,tam2022charge,krieger2022charge,rossi2022broken,tranquada1995evidence,tranquada1995evidence} and nematic states \cite{vojta2009lattice} have been proposed, and are often considered to be related to the superconductivity. Our DFT + U simulations with reasonable Hubbard U and static correction cannot show such evidences. Quite large U value (U = 5 eV) is possible to stabilize the charge density state \cite{chen2022charge}, while such high U value significantly overestimates the Ni-O-Ni rotation angle, strongly underestimates the first-neighbor exchange constant (see Fig. S6) \cite {{note1}} and gives poor agreement with experimental measured excitation spectra (see Fig. S7) \cite {{note1}}. In contrast, despite two competing and possible magnetic ground states have been revealed in the undoped state, the breaking of the commensurate magnetic order from our prediction seems to be intrinsic and robust, and is not affected by the selected Hubbard U, the magnetic ground state in undoped NdNiO$_2$ (C-AFM or 2D-AFM), and the rotation distortion (see Fig. S3) \cite {{note1}}.

In summary, the magnetic properties of infinite-layer nickelates are systematically investigated. We have identified two possible and competing 2D-AFM and 3D AFM ground states for NdNiO$_2$ that can both accurately reproduce recent excitation spectra measurements, which indicates that the current experiments are not sufficient to distinguish the 2D and 3D magnetic characteristics. The microscopic electronic origin of the two possible magnetic states is rationalized through deeper insight into the exchange interactions and the actual orbital contributions. More importantly, hole hoping is found to strongly affect the magnetization of the $d_{x^2-y^2}$ band, magnetic coupling strength, and magnetic order, which naturally trigger a transition to the IC spin order, making then NdNiO$_2$ a true analog of high-T$_c$ cuprates. This is robust and independent on the original ground state of the undoped system. A consistent model connecting the magnetization, exchange interaction, and magnetic order is established and the hole doping tuned nature (AFM or FM) and ratio of first-neighbor and third-neighbor magnetic interactions are revealed to be the decisive factors for the magnetic ground state. 


Y.J.Z. acknowledge financial support by the Initial Scientific Research Fund of Lanzhou University for Young Researcher Fellow (Grant No. 561120206) and the National Natural Science Foundation of China (Grant No. 051B22001). Computations have been performed at the Center for Computational Science and Engineering of Lanzhou University. X. H acknowledges financial support from F.R.S.-FNRS through the PDR Grants PROMOSPAN (T.0107.20). Y.J.Z and X. H contributed equally to this work.

\nocite{*}
\bibliography{NNO-ref}

\begin{thebibliography}{56}%
\makeatletter
\providecommand \@ifxundefined [1]{%
 \@ifx{#1\undefined}
}%
\providecommand \@ifnum [1]{%
 \ifnum #1\expandafter \@firstoftwo
 \else \expandafter \@secondoftwo
 \fi
}%
\providecommand \@ifx [1]{%
 \ifx #1\expandafter \@firstoftwo
 \else \expandafter \@secondoftwo
 \fi
}%
\providecommand \natexlab [1]{#1}%
\providecommand \enquote  [1]{``#1''}%
\providecommand \bibnamefont  [1]{#1}%
\providecommand \bibfnamefont [1]{#1}%
\providecommand \citenamefont [1]{#1}%
\providecommand \href@noop [0]{\@secondoftwo}%
\providecommand \href [0]{\begingroup \@sanitize@url \@href}%
\providecommand \@href[1]{\@@startlink{#1}\@@href}%
\providecommand \@@href[1]{\endgroup#1\@@endlink}%
\providecommand \@sanitize@url [0]{\catcode `\\12\catcode `\$12\catcode
  `\&12\catcode `\#12\catcode `\^12\catcode `\_12\catcode `\%12\relax}%
\providecommand \@@startlink[1]{}%
\providecommand \@@endlink[0]{}%
\providecommand \url  [0]{\begingroup\@sanitize@url \@url }%
\providecommand \@url [1]{\endgroup\@href {#1}{\urlprefix }}%
\providecommand \urlprefix  [0]{URL }%
\providecommand \Eprint [0]{\href }%
\providecommand \doibase [0]{https://doi.org/}%
\providecommand \selectlanguage [0]{\@gobble}%
\providecommand \bibinfo  [0]{\@secondoftwo}%
\providecommand \bibfield  [0]{\@secondoftwo}%
\providecommand \translation [1]{[#1]}%
\providecommand \BibitemOpen [0]{}%
\providecommand \bibitemStop [0]{}%
\providecommand \bibitemNoStop [0]{.\EOS\space}%
\providecommand \EOS [0]{\spacefactor3000\relax}%
\providecommand \BibitemShut  [1]{\csname bibitem#1\endcsname}%
\let\auto@bib@innerbib\@empty
\bibitem [{\citenamefont {Li}\ \emph {et~al.}(2019)\citenamefont {Li},
  \citenamefont {Lee}, \citenamefont {Wang}, \citenamefont {Osada},
  \citenamefont {Crossley}, \citenamefont {Lee}, \citenamefont {Cui},
  \citenamefont {Hikita},\ and\ \citenamefont
  {Hwang}}]{li2019superconductivity}%
  \BibitemOpen
  \bibfield  {author} {\bibinfo {author} {\bibfnamefont {D.}~\bibnamefont
  {Li}}, \bibinfo {author} {\bibfnamefont {K.}~\bibnamefont {Lee}}, \bibinfo
  {author} {\bibfnamefont {B.~Y.}\ \bibnamefont {Wang}}, \bibinfo {author}
  {\bibfnamefont {M.}~\bibnamefont {Osada}}, \bibinfo {author} {\bibfnamefont
  {S.}~\bibnamefont {Crossley}}, \bibinfo {author} {\bibfnamefont {H.~R.}\
  \bibnamefont {Lee}}, \bibinfo {author} {\bibfnamefont {Y.}~\bibnamefont
  {Cui}}, \bibinfo {author} {\bibfnamefont {Y.}~\bibnamefont {Hikita}},\ and\
  \bibinfo {author} {\bibfnamefont {H.~Y.}\ \bibnamefont {Hwang}},\ }\href@noop
  {} {\bibfield  {journal} {\bibinfo  {journal} {Nature}\ }\textbf {\bibinfo
  {volume} {572}},\ \bibinfo {pages} {624} (\bibinfo {year}
  {2019})}\BibitemShut {NoStop}%
\bibitem [{\citenamefont {Osada}\ \emph {et~al.}(2020)\citenamefont {Osada},
  \citenamefont {Wang}, \citenamefont {Goodge}, \citenamefont {Lee},
  \citenamefont {Yoon}, \citenamefont {Sakuma}, \citenamefont {Li},
  \citenamefont {Miura}, \citenamefont {Kourkoutis},\ and\ \citenamefont
  {Hwang}}]{osada2020superconducting}%
  \BibitemOpen
  \bibfield  {author} {\bibinfo {author} {\bibfnamefont {M.}~\bibnamefont
  {Osada}}, \bibinfo {author} {\bibfnamefont {B.~Y.}\ \bibnamefont {Wang}},
  \bibinfo {author} {\bibfnamefont {B.~H.}\ \bibnamefont {Goodge}}, \bibinfo
  {author} {\bibfnamefont {K.}~\bibnamefont {Lee}}, \bibinfo {author}
  {\bibfnamefont {H.}~\bibnamefont {Yoon}}, \bibinfo {author} {\bibfnamefont
  {K.}~\bibnamefont {Sakuma}}, \bibinfo {author} {\bibfnamefont
  {D.}~\bibnamefont {Li}}, \bibinfo {author} {\bibfnamefont {M.}~\bibnamefont
  {Miura}}, \bibinfo {author} {\bibfnamefont {L.~F.}\ \bibnamefont
  {Kourkoutis}},\ and\ \bibinfo {author} {\bibfnamefont {H.~Y.}\ \bibnamefont
  {Hwang}},\ }\href@noop {} {\bibfield  {journal} {\bibinfo  {journal} {Nano
  Lett.}\ }\textbf {\bibinfo {volume} {20}},\ \bibinfo {pages} {5735} (\bibinfo
  {year} {2020})}\BibitemShut {NoStop}%
\bibitem [{\citenamefont {Hepting}\ \emph {et~al.}(2020)\citenamefont
  {Hepting}, \citenamefont {Li}, \citenamefont {Jia}, \citenamefont {Lu},
  \citenamefont {Paris}, \citenamefont {Tseng}, \citenamefont {Feng},
  \citenamefont {Osada}, \citenamefont {Been}, \citenamefont {Hikita} \emph
  {et~al.}}]{hepting2020electronic}%
  \BibitemOpen
  \bibfield  {author} {\bibinfo {author} {\bibfnamefont {M.}~\bibnamefont
  {Hepting}}, \bibinfo {author} {\bibfnamefont {D.}~\bibnamefont {Li}},
  \bibinfo {author} {\bibfnamefont {C.}~\bibnamefont {Jia}}, \bibinfo {author}
  {\bibfnamefont {H.}~\bibnamefont {Lu}}, \bibinfo {author} {\bibfnamefont
  {E.}~\bibnamefont {Paris}}, \bibinfo {author} {\bibfnamefont
  {Y.}~\bibnamefont {Tseng}}, \bibinfo {author} {\bibfnamefont
  {X.}~\bibnamefont {Feng}}, \bibinfo {author} {\bibfnamefont {M.}~\bibnamefont
  {Osada}}, \bibinfo {author} {\bibfnamefont {E.}~\bibnamefont {Been}},
  \bibinfo {author} {\bibfnamefont {Y.}~\bibnamefont {Hikita}}, \emph
  {et~al.},\ }\href@noop {} {\bibfield  {journal} {\bibinfo  {journal} {Nat.
  Mater.}\ }\textbf {\bibinfo {volume} {19}},\ \bibinfo {pages} {381} (\bibinfo
  {year} {2020})}\BibitemShut {NoStop}%
\bibitem [{\citenamefont {Lu}\ \emph {et~al.}(2021)\citenamefont {Lu},
  \citenamefont {Rossi}, \citenamefont {Nag}, \citenamefont {Osada},
  \citenamefont {Li}, \citenamefont {Lee}, \citenamefont {Wang}, \citenamefont
  {Garcia-Fernandez}, \citenamefont {Agrestini}, \citenamefont {Shen},
  \citenamefont {Been}, \citenamefont {Moritz}, \citenamefont {Devereaux},
  \citenamefont {Zaanen}, \citenamefont {Hwang}, \citenamefont {Zhou},\ and\
  \citenamefont {Lee}}]{lu2021magnetic}%
  \BibitemOpen
  \bibfield  {author} {\bibinfo {author} {\bibfnamefont {H.}~\bibnamefont
  {Lu}}, \bibinfo {author} {\bibfnamefont {M.}~\bibnamefont {Rossi}}, \bibinfo
  {author} {\bibfnamefont {A.}~\bibnamefont {Nag}}, \bibinfo {author}
  {\bibfnamefont {M.}~\bibnamefont {Osada}}, \bibinfo {author} {\bibfnamefont
  {D.~F.}\ \bibnamefont {Li}}, \bibinfo {author} {\bibfnamefont
  {K.}~\bibnamefont {Lee}}, \bibinfo {author} {\bibfnamefont {B.~Y.}\
  \bibnamefont {Wang}}, \bibinfo {author} {\bibfnamefont {M.}~\bibnamefont
  {Garcia-Fernandez}}, \bibinfo {author} {\bibfnamefont {S.}~\bibnamefont
  {Agrestini}}, \bibinfo {author} {\bibfnamefont {Z.~X.}\ \bibnamefont {Shen}},
  \bibinfo {author} {\bibfnamefont {E.~M.}\ \bibnamefont {Been}}, \bibinfo
  {author} {\bibfnamefont {B.}~\bibnamefont {Moritz}}, \bibinfo {author}
  {\bibfnamefont {T.~P.}\ \bibnamefont {Devereaux}}, \bibinfo {author}
  {\bibfnamefont {J.}~\bibnamefont {Zaanen}}, \bibinfo {author} {\bibfnamefont
  {H.~Y.}\ \bibnamefont {Hwang}}, \bibinfo {author} {\bibfnamefont {K.-J.}\
  \bibnamefont {Zhou}},\ and\ \bibinfo {author} {\bibfnamefont {W.~S.}\
  \bibnamefont {Lee}},\ }\href@noop {} {\bibfield  {journal} {\bibinfo
  {journal} {Science}\ }\textbf {\bibinfo {volume} {373}},\ \bibinfo {pages}
  {213} (\bibinfo {year} {2021})}\BibitemShut {NoStop}%
\bibitem [{\citenamefont {Wang}\ \emph {et~al.}(2021)\citenamefont {Wang},
  \citenamefont {Li}, \citenamefont {Goodge}, \citenamefont {Lee},
  \citenamefont {Osada}, \citenamefont {Harvey}, \citenamefont {Kourkoutis},
  \citenamefont {Beasley},\ and\ \citenamefont {Hwang}}]{wang2021isotropic}%
  \BibitemOpen
  \bibfield  {author} {\bibinfo {author} {\bibfnamefont {B.~Y.}\ \bibnamefont
  {Wang}}, \bibinfo {author} {\bibfnamefont {D.}~\bibnamefont {Li}}, \bibinfo
  {author} {\bibfnamefont {B.~H.}\ \bibnamefont {Goodge}}, \bibinfo {author}
  {\bibfnamefont {K.}~\bibnamefont {Lee}}, \bibinfo {author} {\bibfnamefont
  {M.}~\bibnamefont {Osada}}, \bibinfo {author} {\bibfnamefont {S.~P.}\
  \bibnamefont {Harvey}}, \bibinfo {author} {\bibfnamefont {L.~F.}\
  \bibnamefont {Kourkoutis}}, \bibinfo {author} {\bibfnamefont {M.~R.}\
  \bibnamefont {Beasley}},\ and\ \bibinfo {author} {\bibfnamefont {H.~Y.}\
  \bibnamefont {Hwang}},\ }\href@noop {} {\bibfield  {journal} {\bibinfo
  {journal} {Nat. Phys.}\ }\textbf {\bibinfo {volume} {17}},\ \bibinfo {pages}
  {473} (\bibinfo {year} {2021})}\BibitemShut {NoStop}%
\bibitem [{\citenamefont {Leonov}\ \emph {et~al.}(2020)\citenamefont {Leonov},
  \citenamefont {Skornyakov},\ and\ \citenamefont
  {Savrasov}}]{leonov2020lifshitz}%
  \BibitemOpen
  \bibfield  {author} {\bibinfo {author} {\bibfnamefont {I.}~\bibnamefont
  {Leonov}}, \bibinfo {author} {\bibfnamefont {S.}~\bibnamefont {Skornyakov}},\
  and\ \bibinfo {author} {\bibfnamefont {S.}~\bibnamefont {Savrasov}},\
  }\href@noop {} {\bibfield  {journal} {\bibinfo  {journal} {Phys. Rev. B}\
  }\textbf {\bibinfo {volume} {101}},\ \bibinfo {pages} {241108} (\bibinfo
  {year} {2020})}\BibitemShut {NoStop}%
\bibitem [{\citenamefont {Ryee}\ \emph {et~al.}(2020)\citenamefont {Ryee},
  \citenamefont {Yoon}, \citenamefont {Kim}, \citenamefont {Jeong},\ and\
  \citenamefont {Han}}]{ryee2020induced}%
  \BibitemOpen
  \bibfield  {author} {\bibinfo {author} {\bibfnamefont {S.}~\bibnamefont
  {Ryee}}, \bibinfo {author} {\bibfnamefont {H.}~\bibnamefont {Yoon}}, \bibinfo
  {author} {\bibfnamefont {T.~J.}\ \bibnamefont {Kim}}, \bibinfo {author}
  {\bibfnamefont {M.~Y.}\ \bibnamefont {Jeong}},\ and\ \bibinfo {author}
  {\bibfnamefont {M.~J.}\ \bibnamefont {Han}},\ }\href@noop {} {\bibfield
  {journal} {\bibinfo  {journal} {Phys. Rev. B}\ }\textbf {\bibinfo {volume}
  {101}},\ \bibinfo {pages} {064513} (\bibinfo {year} {2020})}\BibitemShut
  {NoStop}%
\bibitem [{\citenamefont {Kapeghian}\ and\ \citenamefont
  {Botana}(2020)}]{kapeghian2020electronic}%
  \BibitemOpen
  \bibfield  {author} {\bibinfo {author} {\bibfnamefont {J.}~\bibnamefont
  {Kapeghian}}\ and\ \bibinfo {author} {\bibfnamefont {A.~S.}\ \bibnamefont
  {Botana}},\ }\href@noop {} {\bibfield  {journal} {\bibinfo  {journal} {Phys.
  Rev. B}\ }\textbf {\bibinfo {volume} {102}},\ \bibinfo {pages} {205130}
  (\bibinfo {year} {2020})}\BibitemShut {NoStop}%
\bibitem [{\citenamefont {Botana}\ and\ \citenamefont
  {Norman}(2020)}]{botana2020similarities}%
  \BibitemOpen
  \bibfield  {author} {\bibinfo {author} {\bibfnamefont {A.}~\bibnamefont
  {Botana}}\ and\ \bibinfo {author} {\bibfnamefont {M.}~\bibnamefont
  {Norman}},\ }\href@noop {} {\bibfield  {journal} {\bibinfo  {journal} {Phys.
  Rev. X}\ }\textbf {\bibinfo {volume} {10}},\ \bibinfo {pages} {011024}
  (\bibinfo {year} {2020})}\BibitemShut {NoStop}%
\bibitem [{\citenamefont {Liu}\ \emph {et~al.}(2020)\citenamefont {Liu},
  \citenamefont {Ren}, \citenamefont {Zhu}, \citenamefont {Wang},\ and\
  \citenamefont {Yang}}]{liu2020electronic}%
  \BibitemOpen
  \bibfield  {author} {\bibinfo {author} {\bibfnamefont {Z.}~\bibnamefont
  {Liu}}, \bibinfo {author} {\bibfnamefont {Z.}~\bibnamefont {Ren}}, \bibinfo
  {author} {\bibfnamefont {W.}~\bibnamefont {Zhu}}, \bibinfo {author}
  {\bibfnamefont {Z.}~\bibnamefont {Wang}},\ and\ \bibinfo {author}
  {\bibfnamefont {J.}~\bibnamefont {Yang}},\ }\href@noop {} {\bibfield
  {journal} {\bibinfo  {journal} {npj Quantum Mater.}\ }\textbf {\bibinfo
  {volume} {5}},\ \bibinfo {pages} {1} (\bibinfo {year} {2020})}\BibitemShut
  {NoStop}%
\bibitem [{\citenamefont {Zhang}\ \emph {et~al.}(2021)\citenamefont {Zhang},
  \citenamefont {Lane}, \citenamefont {Singh}, \citenamefont {Nokelainen},
  \citenamefont {Barbiellini}, \citenamefont {Markiewicz}, \citenamefont
  {Bansil},\ and\ \citenamefont {Sun}}]{zhang2021magnetic}%
  \BibitemOpen
  \bibfield  {author} {\bibinfo {author} {\bibfnamefont {R.}~\bibnamefont
  {Zhang}}, \bibinfo {author} {\bibfnamefont {C.}~\bibnamefont {Lane}},
  \bibinfo {author} {\bibfnamefont {B.}~\bibnamefont {Singh}}, \bibinfo
  {author} {\bibfnamefont {J.}~\bibnamefont {Nokelainen}}, \bibinfo {author}
  {\bibfnamefont {B.}~\bibnamefont {Barbiellini}}, \bibinfo {author}
  {\bibfnamefont {R.~S.}\ \bibnamefont {Markiewicz}}, \bibinfo {author}
  {\bibfnamefont {A.}~\bibnamefont {Bansil}},\ and\ \bibinfo {author}
  {\bibfnamefont {J.}~\bibnamefont {Sun}},\ }\href@noop {} {\bibfield
  {journal} {\bibinfo  {journal} {Commun. Phys.}\ }\textbf {\bibinfo {volume}
  {4}},\ \bibinfo {pages} {1} (\bibinfo {year} {2021})}\BibitemShut {NoStop}%
\bibitem [{\citenamefont {Been}\ \emph {et~al.}(2021)\citenamefont {Been},
  \citenamefont {Lee}, \citenamefont {Hwang}, \citenamefont {Cui},
  \citenamefont {Zaanen}, \citenamefont {Devereaux}, \citenamefont {Moritz},\
  and\ \citenamefont {Jia}}]{been2021electronic}%
  \BibitemOpen
  \bibfield  {author} {\bibinfo {author} {\bibfnamefont {E.}~\bibnamefont
  {Been}}, \bibinfo {author} {\bibfnamefont {W.-S.}\ \bibnamefont {Lee}},
  \bibinfo {author} {\bibfnamefont {H.~Y.}\ \bibnamefont {Hwang}}, \bibinfo
  {author} {\bibfnamefont {Y.}~\bibnamefont {Cui}}, \bibinfo {author}
  {\bibfnamefont {J.}~\bibnamefont {Zaanen}}, \bibinfo {author} {\bibfnamefont
  {T.}~\bibnamefont {Devereaux}}, \bibinfo {author} {\bibfnamefont
  {B.}~\bibnamefont {Moritz}},\ and\ \bibinfo {author} {\bibfnamefont
  {C.}~\bibnamefont {Jia}},\ }\href@noop {} {\bibfield  {journal} {\bibinfo
  {journal} {Phys. Rev. X}\ }\textbf {\bibinfo {volume} {11}},\ \bibinfo
  {pages} {011050} (\bibinfo {year} {2021})}\BibitemShut {NoStop}%
\bibitem [{\citenamefont {Fowlie}\ \emph {et~al.}(2022)\citenamefont {Fowlie},
  \citenamefont {Hadjimichael}, \citenamefont {Martins}, \citenamefont {Li},
  \citenamefont {Osada}, \citenamefont {Wang}, \citenamefont {Lee},
  \citenamefont {Lee}, \citenamefont {Salman}, \citenamefont {Prokscha} \emph
  {et~al.}}]{fowlie2022intrinsic}%
  \BibitemOpen
  \bibfield  {author} {\bibinfo {author} {\bibfnamefont {J.}~\bibnamefont
  {Fowlie}}, \bibinfo {author} {\bibfnamefont {M.}~\bibnamefont
  {Hadjimichael}}, \bibinfo {author} {\bibfnamefont {M.~M.}\ \bibnamefont
  {Martins}}, \bibinfo {author} {\bibfnamefont {D.}~\bibnamefont {Li}},
  \bibinfo {author} {\bibfnamefont {M.}~\bibnamefont {Osada}}, \bibinfo
  {author} {\bibfnamefont {B.~Y.}\ \bibnamefont {Wang}}, \bibinfo {author}
  {\bibfnamefont {K.}~\bibnamefont {Lee}}, \bibinfo {author} {\bibfnamefont
  {Y.}~\bibnamefont {Lee}}, \bibinfo {author} {\bibfnamefont {Z.}~\bibnamefont
  {Salman}}, \bibinfo {author} {\bibfnamefont {T.}~\bibnamefont {Prokscha}},
  \emph {et~al.},\ }\href@noop {} {\bibfield  {journal} {\bibinfo  {journal}
  {arXiv preprint arXiv:2201.11943}\ } (\bibinfo {year} {2022})}\BibitemShut
  {NoStop}%
\bibitem [{\citenamefont {Jiang}\ \emph {et~al.}(2020)\citenamefont {Jiang},
  \citenamefont {Berciu},\ and\ \citenamefont {Sawatzky}}]{jiang2020critical}%
  \BibitemOpen
  \bibfield  {author} {\bibinfo {author} {\bibfnamefont {M.}~\bibnamefont
  {Jiang}}, \bibinfo {author} {\bibfnamefont {M.}~\bibnamefont {Berciu}},\ and\
  \bibinfo {author} {\bibfnamefont {G.~A.}\ \bibnamefont {Sawatzky}},\
  }\href@noop {} {\bibfield  {journal} {\bibinfo  {journal} {Phys. Rev. Lett.}\
  }\textbf {\bibinfo {volume} {124}},\ \bibinfo {pages} {207004} (\bibinfo
  {year} {2020})}\BibitemShut {NoStop}%
\bibitem [{\citenamefont {Wan}\ \emph {et~al.}(2021)\citenamefont {Wan},
  \citenamefont {Ivanov}, \citenamefont {Resta}, \citenamefont {Leonov},\ and\
  \citenamefont {Savrasov}}]{wan2021exchange}%
  \BibitemOpen
  \bibfield  {author} {\bibinfo {author} {\bibfnamefont {X.}~\bibnamefont
  {Wan}}, \bibinfo {author} {\bibfnamefont {V.}~\bibnamefont {Ivanov}},
  \bibinfo {author} {\bibfnamefont {G.}~\bibnamefont {Resta}}, \bibinfo
  {author} {\bibfnamefont {I.}~\bibnamefont {Leonov}},\ and\ \bibinfo {author}
  {\bibfnamefont {S.~Y.}\ \bibnamefont {Savrasov}},\ }\href@noop {} {\bibfield
  {journal} {\bibinfo  {journal} {Phys. Rev. B}\ }\textbf {\bibinfo {volume}
  {103}},\ \bibinfo {pages} {075123} (\bibinfo {year} {2021})}\BibitemShut
  {NoStop}%
\bibitem [{\citenamefont {Li}\ \emph {et~al.}(2020)\citenamefont {Li},
  \citenamefont {Wang}, \citenamefont {Lee}, \citenamefont {Harvey},
  \citenamefont {Osada}, \citenamefont {Goodge}, \citenamefont {Kourkoutis},\
  and\ \citenamefont {Hwang}}]{li2020superconducting}%
  \BibitemOpen
  \bibfield  {author} {\bibinfo {author} {\bibfnamefont {D.}~\bibnamefont
  {Li}}, \bibinfo {author} {\bibfnamefont {B.~Y.}\ \bibnamefont {Wang}},
  \bibinfo {author} {\bibfnamefont {K.}~\bibnamefont {Lee}}, \bibinfo {author}
  {\bibfnamefont {S.~P.}\ \bibnamefont {Harvey}}, \bibinfo {author}
  {\bibfnamefont {M.}~\bibnamefont {Osada}}, \bibinfo {author} {\bibfnamefont
  {B.~H.}\ \bibnamefont {Goodge}}, \bibinfo {author} {\bibfnamefont {L.~F.}\
  \bibnamefont {Kourkoutis}},\ and\ \bibinfo {author} {\bibfnamefont {H.~Y.}\
  \bibnamefont {Hwang}},\ }\href@noop {} {\bibfield  {journal} {\bibinfo
  {journal} {Phys. Rev. Lett.}\ }\textbf {\bibinfo {volume} {125}},\ \bibinfo
  {pages} {027001} (\bibinfo {year} {2020})}\BibitemShut {NoStop}%
\bibitem [{\citenamefont {Goodge}\ \emph {et~al.}(2021)\citenamefont {Goodge},
  \citenamefont {Li}, \citenamefont {Lee}, \citenamefont {Osada}, \citenamefont
  {Wang}, \citenamefont {Sawatzky}, \citenamefont {Hwang},\ and\ \citenamefont
  {Kourkoutis}}]{goodge2021doping}%
  \BibitemOpen
  \bibfield  {author} {\bibinfo {author} {\bibfnamefont {B.~H.}\ \bibnamefont
  {Goodge}}, \bibinfo {author} {\bibfnamefont {D.}~\bibnamefont {Li}}, \bibinfo
  {author} {\bibfnamefont {K.}~\bibnamefont {Lee}}, \bibinfo {author}
  {\bibfnamefont {M.}~\bibnamefont {Osada}}, \bibinfo {author} {\bibfnamefont
  {B.~Y.}\ \bibnamefont {Wang}}, \bibinfo {author} {\bibfnamefont {G.~A.}\
  \bibnamefont {Sawatzky}}, \bibinfo {author} {\bibfnamefont {H.~Y.}\
  \bibnamefont {Hwang}},\ and\ \bibinfo {author} {\bibfnamefont {L.~F.}\
  \bibnamefont {Kourkoutis}},\ }\href@noop {} {\bibfield  {journal} {\bibinfo
  {journal} {Proc. Natl. Acad. Sci. U.S.A.}\ }\textbf {\bibinfo {volume} {118}}
  (\bibinfo {year} {2021})}\BibitemShut {NoStop}%
\bibitem [{\citenamefont {Chen}\ \emph
  {et~al.}(2022{\natexlab{a}})\citenamefont {Chen}, \citenamefont {Jiang},
  \citenamefont {Si}, \citenamefont {Lu},\ and\ \citenamefont
  {Zhong}}]{chen2022magnetism}%
  \BibitemOpen
  \bibfield  {author} {\bibinfo {author} {\bibfnamefont {D.}~\bibnamefont
  {Chen}}, \bibinfo {author} {\bibfnamefont {P.}~\bibnamefont {Jiang}},
  \bibinfo {author} {\bibfnamefont {L.}~\bibnamefont {Si}}, \bibinfo {author}
  {\bibfnamefont {Y.}~\bibnamefont {Lu}},\ and\ \bibinfo {author}
  {\bibfnamefont {Z.}~\bibnamefont {Zhong}},\ }\href@noop {} {\bibfield
  {journal} {\bibinfo  {journal} {Phys. Rev. B}\ }\textbf {\bibinfo {volume}
  {106}},\ \bibinfo {pages} {045105} (\bibinfo {year}
  {2022}{\natexlab{a}})}\BibitemShut {NoStop}%
\bibitem [{\citenamefont {Krishna}\ \emph {et~al.}(2020)\citenamefont
  {Krishna}, \citenamefont {LaBollita}, \citenamefont {Fumega}, \citenamefont
  {Pardo},\ and\ \citenamefont {Botana}}]{krishna2020effects}%
  \BibitemOpen
  \bibfield  {author} {\bibinfo {author} {\bibfnamefont {J.}~\bibnamefont
  {Krishna}}, \bibinfo {author} {\bibfnamefont {H.}~\bibnamefont {LaBollita}},
  \bibinfo {author} {\bibfnamefont {A.~O.}\ \bibnamefont {Fumega}}, \bibinfo
  {author} {\bibfnamefont {V.}~\bibnamefont {Pardo}},\ and\ \bibinfo {author}
  {\bibfnamefont {A.~S.}\ \bibnamefont {Botana}},\ }\href@noop {} {\bibfield
  {journal} {\bibinfo  {journal} {Phys. Rev. B}\ }\textbf {\bibinfo {volume}
  {102}},\ \bibinfo {pages} {224506} (\bibinfo {year} {2020})}\BibitemShut
  {NoStop}%
\bibitem [{\citenamefont {Nomura}\ \emph {et~al.}(2019)\citenamefont {Nomura},
  \citenamefont {Hirayama}, \citenamefont {Tadano}, \citenamefont {Yoshimoto},
  \citenamefont {Nakamura},\ and\ \citenamefont {Arita}}]{nomura2019formation}%
  \BibitemOpen
  \bibfield  {author} {\bibinfo {author} {\bibfnamefont {Y.}~\bibnamefont
  {Nomura}}, \bibinfo {author} {\bibfnamefont {M.}~\bibnamefont {Hirayama}},
  \bibinfo {author} {\bibfnamefont {T.}~\bibnamefont {Tadano}}, \bibinfo
  {author} {\bibfnamefont {Y.}~\bibnamefont {Yoshimoto}}, \bibinfo {author}
  {\bibfnamefont {K.}~\bibnamefont {Nakamura}},\ and\ \bibinfo {author}
  {\bibfnamefont {R.}~\bibnamefont {Arita}},\ }\href@noop {} {\bibfield
  {journal} {\bibinfo  {journal} {Phys. Rev. B}\ }\textbf {\bibinfo {volume}
  {100}},\ \bibinfo {pages} {205138} (\bibinfo {year} {2019})}\BibitemShut
  {NoStop}%
\bibitem [{\citenamefont {Xia}\ \emph {et~al.}(2022)\citenamefont {Xia},
  \citenamefont {Wu}, \citenamefont {Chen},\ and\ \citenamefont
  {Chen}}]{xia2022dynamical}%
  \BibitemOpen
  \bibfield  {author} {\bibinfo {author} {\bibfnamefont {C.}~\bibnamefont
  {Xia}}, \bibinfo {author} {\bibfnamefont {J.}~\bibnamefont {Wu}}, \bibinfo
  {author} {\bibfnamefont {Y.}~\bibnamefont {Chen}},\ and\ \bibinfo {author}
  {\bibfnamefont {H.}~\bibnamefont {Chen}},\ }\href@noop {} {\bibfield
  {journal} {\bibinfo  {journal} {Phys. Rev. B}\ }\textbf {\bibinfo {volume}
  {105}},\ \bibinfo {pages} {115134} (\bibinfo {year} {2022})}\BibitemShut
  {NoStop}%
\bibitem [{\citenamefont {Bernardini}\ \emph {et~al.}(2022)\citenamefont
  {Bernardini}, \citenamefont {Bosin},\ and\ \citenamefont
  {Cano}}]{bernardini2022geometric}%
  \BibitemOpen
  \bibfield  {author} {\bibinfo {author} {\bibfnamefont {F.}~\bibnamefont
  {Bernardini}}, \bibinfo {author} {\bibfnamefont {A.}~\bibnamefont {Bosin}},\
  and\ \bibinfo {author} {\bibfnamefont {A.}~\bibnamefont {Cano}},\ }\href@noop
  {} {\bibfield  {journal} {\bibinfo  {journal} {Phys. Rev. Mater.}\ }\textbf
  {\bibinfo {volume} {6}},\ \bibinfo {pages} {044807} (\bibinfo {year}
  {2022})}\BibitemShut {NoStop}%
\bibitem [{\citenamefont {Tam}\ \emph {et~al.}(2022)\citenamefont {Tam},
  \citenamefont {Choi}, \citenamefont {Ding}, \citenamefont {Agrestini},
  \citenamefont {Nag}, \citenamefont {Wu}, \citenamefont {Huang}, \citenamefont
  {Luo}, \citenamefont {Gao}, \citenamefont {Garc{\'\i}a-Fern{\'a}ndez} \emph
  {et~al.}}]{tam2022charge}%
  \BibitemOpen
  \bibfield  {author} {\bibinfo {author} {\bibfnamefont {C.~C.}\ \bibnamefont
  {Tam}}, \bibinfo {author} {\bibfnamefont {J.}~\bibnamefont {Choi}}, \bibinfo
  {author} {\bibfnamefont {X.}~\bibnamefont {Ding}}, \bibinfo {author}
  {\bibfnamefont {S.}~\bibnamefont {Agrestini}}, \bibinfo {author}
  {\bibfnamefont {A.}~\bibnamefont {Nag}}, \bibinfo {author} {\bibfnamefont
  {M.}~\bibnamefont {Wu}}, \bibinfo {author} {\bibfnamefont {B.}~\bibnamefont
  {Huang}}, \bibinfo {author} {\bibfnamefont {H.}~\bibnamefont {Luo}}, \bibinfo
  {author} {\bibfnamefont {P.}~\bibnamefont {Gao}}, \bibinfo {author}
  {\bibfnamefont {M.}~\bibnamefont {Garc{\'\i}a-Fern{\'a}ndez}}, \emph
  {et~al.},\ }\href@noop {} {\bibfield  {journal} {\bibinfo  {journal} {Nat.
  Mater.}\ ,\ \bibinfo {pages} {1}} (\bibinfo {year} {2022})}\BibitemShut
  {NoStop}%
\bibitem [{\citenamefont {Krieger}\ \emph {et~al.}(2022)\citenamefont
  {Krieger}, \citenamefont {Martinelli}, \citenamefont {Zeng}, \citenamefont
  {Chow}, \citenamefont {Kummer}, \citenamefont {Arpaia}, \citenamefont {Sala},
  \citenamefont {Brookes}, \citenamefont {Ariando}, \citenamefont {Viart} \emph
  {et~al.}}]{krieger2022charge}%
  \BibitemOpen
  \bibfield  {author} {\bibinfo {author} {\bibfnamefont {G.}~\bibnamefont
  {Krieger}}, \bibinfo {author} {\bibfnamefont {L.}~\bibnamefont {Martinelli}},
  \bibinfo {author} {\bibfnamefont {S.}~\bibnamefont {Zeng}}, \bibinfo {author}
  {\bibfnamefont {L.}~\bibnamefont {Chow}}, \bibinfo {author} {\bibfnamefont
  {K.}~\bibnamefont {Kummer}}, \bibinfo {author} {\bibfnamefont
  {R.}~\bibnamefont {Arpaia}}, \bibinfo {author} {\bibfnamefont {M.~M.}\
  \bibnamefont {Sala}}, \bibinfo {author} {\bibfnamefont {N.}~\bibnamefont
  {Brookes}}, \bibinfo {author} {\bibfnamefont {A.}~\bibnamefont {Ariando}},
  \bibinfo {author} {\bibfnamefont {N.}~\bibnamefont {Viart}}, \emph {et~al.},\
  }\href@noop {} {\bibfield  {journal} {\bibinfo  {journal} {Phys. Rev. Lett.}\
  }\textbf {\bibinfo {volume} {129}},\ \bibinfo {pages} {027002} (\bibinfo
  {year} {2022})}\BibitemShut {NoStop}%
\bibitem [{\citenamefont {Rossi}\ \emph {et~al.}(2022)\citenamefont {Rossi},
  \citenamefont {Osada}, \citenamefont {Choi}, \citenamefont {Agrestini},
  \citenamefont {Jost}, \citenamefont {Lee}, \citenamefont {Lu}, \citenamefont
  {Wang}, \citenamefont {Lee}, \citenamefont {Nag} \emph
  {et~al.}}]{rossi2022broken}%
  \BibitemOpen
  \bibfield  {author} {\bibinfo {author} {\bibfnamefont {M.}~\bibnamefont
  {Rossi}}, \bibinfo {author} {\bibfnamefont {M.}~\bibnamefont {Osada}},
  \bibinfo {author} {\bibfnamefont {J.}~\bibnamefont {Choi}}, \bibinfo {author}
  {\bibfnamefont {S.}~\bibnamefont {Agrestini}}, \bibinfo {author}
  {\bibfnamefont {D.}~\bibnamefont {Jost}}, \bibinfo {author} {\bibfnamefont
  {Y.}~\bibnamefont {Lee}}, \bibinfo {author} {\bibfnamefont {H.}~\bibnamefont
  {Lu}}, \bibinfo {author} {\bibfnamefont {B.~Y.}\ \bibnamefont {Wang}},
  \bibinfo {author} {\bibfnamefont {K.}~\bibnamefont {Lee}}, \bibinfo {author}
  {\bibfnamefont {A.}~\bibnamefont {Nag}}, \emph {et~al.},\ }\href@noop {}
  {\bibfield  {journal} {\bibinfo  {journal} {Nat. Phys.}\ ,\ \bibinfo {pages}
  {1}} (\bibinfo {year} {2022})}\BibitemShut {NoStop}%
\bibitem [{\citenamefont {Chen}\ \emph
  {et~al.}(2022{\natexlab{b}})\citenamefont {Chen}, \citenamefont {Yang},\ and\
  \citenamefont {Zhang}}]{chen2022charge}%
  \BibitemOpen
  \bibfield  {author} {\bibinfo {author} {\bibfnamefont {H.}~\bibnamefont
  {Chen}}, \bibinfo {author} {\bibfnamefont {Y.-f.}\ \bibnamefont {Yang}},\
  and\ \bibinfo {author} {\bibfnamefont {G.-M.}\ \bibnamefont {Zhang}},\
  }\href@noop {} {\bibfield  {journal} {\bibinfo  {journal} {arXiv preprint
  arXiv:2204.12208}\ } (\bibinfo {year} {2022}{\natexlab{b}})}\BibitemShut
  {NoStop}%
\bibitem [{\citenamefont {Moriya}\ and\ \citenamefont
  {Ueda}(2000)}]{moriya2000spin}%
  \BibitemOpen
  \bibfield  {author} {\bibinfo {author} {\bibfnamefont {T.}~\bibnamefont
  {Moriya}}\ and\ \bibinfo {author} {\bibfnamefont {K.}~\bibnamefont {Ueda}},\
  }\href@noop {} {\bibfield  {journal} {\bibinfo  {journal} {Adv. Phys.}\
  }\textbf {\bibinfo {volume} {49}},\ \bibinfo {pages} {555} (\bibinfo {year}
  {2000})}\BibitemShut {NoStop}%
\bibitem [{\citenamefont {Kastner}\ \emph {et~al.}(1998)\citenamefont
  {Kastner}, \citenamefont {Birgeneau}, \citenamefont {Shirane},\ and\
  \citenamefont {Endoh}}]{kastner1998magnetic}%
  \BibitemOpen
  \bibfield  {author} {\bibinfo {author} {\bibfnamefont {M.}~\bibnamefont
  {Kastner}}, \bibinfo {author} {\bibfnamefont {R.}~\bibnamefont {Birgeneau}},
  \bibinfo {author} {\bibfnamefont {G.}~\bibnamefont {Shirane}},\ and\ \bibinfo
  {author} {\bibfnamefont {Y.}~\bibnamefont {Endoh}},\ }\href@noop {}
  {\bibfield  {journal} {\bibinfo  {journal} {Rev. Mod. Phys.}\ }\textbf
  {\bibinfo {volume} {70}},\ \bibinfo {pages} {897} (\bibinfo {year}
  {1998})}\BibitemShut {NoStop}%
\bibitem [{\citenamefont {J.~Birgeneau}\ \emph {et~al.}(2006)\citenamefont
  {J.~Birgeneau}, \citenamefont {Stock}, \citenamefont {M.~Tranquada},\ and\
  \citenamefont {Yamada}}]{j2006magnetic}%
  \BibitemOpen
  \bibfield  {author} {\bibinfo {author} {\bibfnamefont {R.}~\bibnamefont
  {J.~Birgeneau}}, \bibinfo {author} {\bibfnamefont {C.}~\bibnamefont {Stock}},
  \bibinfo {author} {\bibfnamefont {J.}~\bibnamefont {M.~Tranquada}},\ and\
  \bibinfo {author} {\bibfnamefont {K.}~\bibnamefont {Yamada}},\ }\href@noop {}
  {\bibfield  {journal} {\bibinfo  {journal} {J. Phys. Soc. Japan}\ }\textbf
  {\bibinfo {volume} {75}},\ \bibinfo {pages} {111003} (\bibinfo {year}
  {2006})}\BibitemShut {NoStop}%
\bibitem [{\citenamefont {Endoh}\ \emph {et~al.}(1988)\citenamefont {Endoh},
  \citenamefont {Yamada}, \citenamefont {Birgeneau}, \citenamefont {Gabbe},
  \citenamefont {Jenssen}, \citenamefont {Kastner}, \citenamefont {Peters},
  \citenamefont {Picone}, \citenamefont {Thurston}, \citenamefont {Tranquada}
  \emph {et~al.}}]{endoh1988static}%
  \BibitemOpen
  \bibfield  {author} {\bibinfo {author} {\bibfnamefont {Y.}~\bibnamefont
  {Endoh}}, \bibinfo {author} {\bibfnamefont {K.}~\bibnamefont {Yamada}},
  \bibinfo {author} {\bibfnamefont {R.}~\bibnamefont {Birgeneau}}, \bibinfo
  {author} {\bibfnamefont {D.}~\bibnamefont {Gabbe}}, \bibinfo {author}
  {\bibfnamefont {H.}~\bibnamefont {Jenssen}}, \bibinfo {author} {\bibfnamefont
  {M.}~\bibnamefont {Kastner}}, \bibinfo {author} {\bibfnamefont
  {C.}~\bibnamefont {Peters}}, \bibinfo {author} {\bibfnamefont
  {P.}~\bibnamefont {Picone}}, \bibinfo {author} {\bibfnamefont
  {T.}~\bibnamefont {Thurston}}, \bibinfo {author} {\bibfnamefont
  {J.}~\bibnamefont {Tranquada}}, \emph {et~al.},\ }\href@noop {} {\bibfield
  {journal} {\bibinfo  {journal} {Phys. Rev. B}\ }\textbf {\bibinfo {volume}
  {37}},\ \bibinfo {pages} {7443} (\bibinfo {year} {1988})}\BibitemShut
  {NoStop}%
\bibitem [{\citenamefont {Zhao}\ \emph {et~al.}(2009)\citenamefont {Zhao},
  \citenamefont {Adroja}, \citenamefont {Yao}, \citenamefont {Bewley},
  \citenamefont {Li}, \citenamefont {Wang}, \citenamefont {Wu}, \citenamefont
  {Chen}, \citenamefont {Hu},\ and\ \citenamefont {Dai}}]{zhao2009spin}%
  \BibitemOpen
  \bibfield  {author} {\bibinfo {author} {\bibfnamefont {J.}~\bibnamefont
  {Zhao}}, \bibinfo {author} {\bibfnamefont {D.}~\bibnamefont {Adroja}},
  \bibinfo {author} {\bibfnamefont {D.-X.}\ \bibnamefont {Yao}}, \bibinfo
  {author} {\bibfnamefont {R.}~\bibnamefont {Bewley}}, \bibinfo {author}
  {\bibfnamefont {S.}~\bibnamefont {Li}}, \bibinfo {author} {\bibfnamefont
  {X.}~\bibnamefont {Wang}}, \bibinfo {author} {\bibfnamefont {G.}~\bibnamefont
  {Wu}}, \bibinfo {author} {\bibfnamefont {X.}~\bibnamefont {Chen}}, \bibinfo
  {author} {\bibfnamefont {J.}~\bibnamefont {Hu}},\ and\ \bibinfo {author}
  {\bibfnamefont {P.}~\bibnamefont {Dai}},\ }\href@noop {} {\bibfield
  {journal} {\bibinfo  {journal} {Nat. Phys.}\ }\textbf {\bibinfo {volume}
  {5}},\ \bibinfo {pages} {555} (\bibinfo {year} {2009})}\BibitemShut {NoStop}%
\bibitem [{\citenamefont {Perdew}\ \emph {et~al.}(1996)\citenamefont {Perdew},
  \citenamefont {Burke},\ and\ \citenamefont
  {Ernzerhof}}]{perdew1996generalized}%
  \BibitemOpen
  \bibfield  {author} {\bibinfo {author} {\bibfnamefont {J.~P.}\ \bibnamefont
  {Perdew}}, \bibinfo {author} {\bibfnamefont {K.}~\bibnamefont {Burke}},\ and\
  \bibinfo {author} {\bibfnamefont {M.}~\bibnamefont {Ernzerhof}},\ }\href@noop
  {} {\bibfield  {journal} {\bibinfo  {journal} {Phys. Rev. Lett.}\ }\textbf
  {\bibinfo {volume} {77}},\ \bibinfo {pages} {3865} (\bibinfo {year}
  {1996})}\BibitemShut {NoStop}%
\bibitem [{\citenamefont {Dudarev}\ \emph {et~al.}(1998)\citenamefont
  {Dudarev}, \citenamefont {Botton}, \citenamefont {Savrasov}, \citenamefont
  {Humphreys},\ and\ \citenamefont {Sutton}}]{dudarev1998electron}%
  \BibitemOpen
  \bibfield  {author} {\bibinfo {author} {\bibfnamefont {S.}~\bibnamefont
  {Dudarev}}, \bibinfo {author} {\bibfnamefont {G.}~\bibnamefont {Botton}},
  \bibinfo {author} {\bibfnamefont {S.}~\bibnamefont {Savrasov}}, \bibinfo
  {author} {\bibfnamefont {C.}~\bibnamefont {Humphreys}},\ and\ \bibinfo
  {author} {\bibfnamefont {A.}~\bibnamefont {Sutton}},\ }\href@noop {}
  {\bibfield  {journal} {\bibinfo  {journal} {Phys. Rev. B}\ }\textbf {\bibinfo
  {volume} {57}},\ \bibinfo {pages} {1505} (\bibinfo {year}
  {1998})}\BibitemShut {NoStop}%
\bibitem [{not()}]{note1}%
  \BibitemOpen
  \href@noop {} {}\bibinfo {note} {See Supplemental Material, which includes
  Refs. [35-46], the (i) computational details; (ii) ground state properties of
  $R$NiO$_2$; (iii) exchange constants and the orbital contributions; (iv)
  doping concentration dependent incommensurability $\delta$ under different
  conditions; (vi) Hubbard U dependent spin-wave dispersion.}\BibitemShut
  {Stop}%
\bibitem [{\citenamefont {He}\ \emph {et~al.}(2021)\citenamefont {He},
  \citenamefont {Helbig}, \citenamefont {Verstraete},\ and\ \citenamefont
  {Bousquet}}]{he2021tb2j}%
  \BibitemOpen
  \bibfield  {author} {\bibinfo {author} {\bibfnamefont {X.}~\bibnamefont
  {He}}, \bibinfo {author} {\bibfnamefont {N.}~\bibnamefont {Helbig}}, \bibinfo
  {author} {\bibfnamefont {M.~J.}\ \bibnamefont {Verstraete}},\ and\ \bibinfo
  {author} {\bibfnamefont {E.}~\bibnamefont {Bousquet}},\ }\href@noop {}
  {\bibfield  {journal} {\bibinfo  {journal} {Comput. Phys. Commun.}\ }\textbf
  {\bibinfo {volume} {264}},\ \bibinfo {pages} {107938} (\bibinfo {year}
  {2021})}\BibitemShut {NoStop}%
\bibitem [{\citenamefont {Toth}\ and\ \citenamefont
  {Lake}(2015)}]{toth2015linear}%
  \BibitemOpen
  \bibfield  {author} {\bibinfo {author} {\bibfnamefont {S.}~\bibnamefont
  {Toth}}\ and\ \bibinfo {author} {\bibfnamefont {B.}~\bibnamefont {Lake}},\
  }\href@noop {} {\bibfield  {journal} {\bibinfo  {journal} {J. Phys.: Condens.
  Matter}\ }\textbf {\bibinfo {volume} {27}},\ \bibinfo {pages} {166002}
  (\bibinfo {year} {2015})}\BibitemShut {NoStop}%
\bibitem [{\citenamefont {Kresse}\ and\ \citenamefont
  {Hafner}(1993)}]{kresse1993ab}%
  \BibitemOpen
  \bibfield  {author} {\bibinfo {author} {\bibfnamefont {G.}~\bibnamefont
  {Kresse}}\ and\ \bibinfo {author} {\bibfnamefont {J.}~\bibnamefont
  {Hafner}},\ }\href@noop {} {\bibfield  {journal} {\bibinfo  {journal} {Phys.
  Rev. B}\ }\textbf {\bibinfo {volume} {47}},\ \bibinfo {pages} {558} (\bibinfo
  {year} {1993})}\BibitemShut {NoStop}%
\bibitem [{\citenamefont {Bl{\"o}chl}(1994)}]{blochl1994projector}%
  \BibitemOpen
  \bibfield  {author} {\bibinfo {author} {\bibfnamefont {P.~E.}\ \bibnamefont
  {Bl{\"o}chl}},\ }\href@noop {} {\bibfield  {journal} {\bibinfo  {journal}
  {Phys. Rev. B}\ }\textbf {\bibinfo {volume} {50}},\ \bibinfo {pages} {17953}
  (\bibinfo {year} {1994})}\BibitemShut {NoStop}%
\bibitem [{\citenamefont {Sun}\ \emph {et~al.}(2015)\citenamefont {Sun},
  \citenamefont {Ruzsinszky},\ and\ \citenamefont {Perdew}}]{sun2015strongly}%
  \BibitemOpen
  \bibfield  {author} {\bibinfo {author} {\bibfnamefont {J.}~\bibnamefont
  {Sun}}, \bibinfo {author} {\bibfnamefont {A.}~\bibnamefont {Ruzsinszky}},\
  and\ \bibinfo {author} {\bibfnamefont {J.~P.}\ \bibnamefont {Perdew}},\
  }\href@noop {} {\bibfield  {journal} {\bibinfo  {journal} {Phys. Rev. Lett.}\
  }\textbf {\bibinfo {volume} {115}},\ \bibinfo {pages} {036402} (\bibinfo
  {year} {2015})}\BibitemShut {NoStop}%
\bibitem [{\citenamefont {Fu}\ and\ \citenamefont
  {Singh}(2018)}]{fu2018applicability}%
  \BibitemOpen
  \bibfield  {author} {\bibinfo {author} {\bibfnamefont {Y.}~\bibnamefont
  {Fu}}\ and\ \bibinfo {author} {\bibfnamefont {D.~J.}\ \bibnamefont {Singh}},\
  }\href@noop {} {\bibfield  {journal} {\bibinfo  {journal} {Phys. Rev. Lett.}\
  }\textbf {\bibinfo {volume} {121}},\ \bibinfo {pages} {207201} (\bibinfo
  {year} {2018})}\BibitemShut {NoStop}%
\bibitem [{\citenamefont {Monkhorst}\ and\ \citenamefont
  {Pack}(1976)}]{monkhorst1976special}%
  \BibitemOpen
  \bibfield  {author} {\bibinfo {author} {\bibfnamefont {H.~J.}\ \bibnamefont
  {Monkhorst}}\ and\ \bibinfo {author} {\bibfnamefont {J.~D.}\ \bibnamefont
  {Pack}},\ }\href@noop {} {\bibfield  {journal} {\bibinfo  {journal} {Phys.
  Rev. B}\ }\textbf {\bibinfo {volume} {13}},\ \bibinfo {pages} {5188}
  (\bibinfo {year} {1976})}\BibitemShut {NoStop}%
\bibitem [{\citenamefont {Togo}\ \emph {et~al.}(2008)\citenamefont {Togo},
  \citenamefont {Oba},\ and\ \citenamefont {Tanaka}}]{togo2008first}%
  \BibitemOpen
  \bibfield  {author} {\bibinfo {author} {\bibfnamefont {A.}~\bibnamefont
  {Togo}}, \bibinfo {author} {\bibfnamefont {F.}~\bibnamefont {Oba}},\ and\
  \bibinfo {author} {\bibfnamefont {I.}~\bibnamefont {Tanaka}},\ }\href@noop {}
  {\bibfield  {journal} {\bibinfo  {journal} {Phys. Rev. B}\ }\textbf {\bibinfo
  {volume} {78}},\ \bibinfo {pages} {134106} (\bibinfo {year}
  {2008})}\BibitemShut {NoStop}%
\bibitem [{\citenamefont {Marzari}\ \emph {et~al.}(2012)\citenamefont
  {Marzari}, \citenamefont {Mostofi}, \citenamefont {Yates}, \citenamefont
  {Souza},\ and\ \citenamefont {Vanderbilt}}]{marzari2012maximally}%
  \BibitemOpen
  \bibfield  {author} {\bibinfo {author} {\bibfnamefont {N.}~\bibnamefont
  {Marzari}}, \bibinfo {author} {\bibfnamefont {A.~A.}\ \bibnamefont
  {Mostofi}}, \bibinfo {author} {\bibfnamefont {J.~R.}\ \bibnamefont {Yates}},
  \bibinfo {author} {\bibfnamefont {I.}~\bibnamefont {Souza}},\ and\ \bibinfo
  {author} {\bibfnamefont {D.}~\bibnamefont {Vanderbilt}},\ }\href@noop {}
  {\bibfield  {journal} {\bibinfo  {journal} {Rev. Mod. Phys.}\ }\textbf
  {\bibinfo {volume} {84}},\ \bibinfo {pages} {1419} (\bibinfo {year}
  {2012})}\BibitemShut {NoStop}%
\bibitem [{\citenamefont {Mostofi}\ \emph {et~al.}(2014)\citenamefont
  {Mostofi}, \citenamefont {Yates}, \citenamefont {Pizzi}, \citenamefont {Lee},
  \citenamefont {Souza}, \citenamefont {Vanderbilt},\ and\ \citenamefont
  {Marzari}}]{mostofi2014updated}%
  \BibitemOpen
  \bibfield  {author} {\bibinfo {author} {\bibfnamefont {A.~A.}\ \bibnamefont
  {Mostofi}}, \bibinfo {author} {\bibfnamefont {J.~R.}\ \bibnamefont {Yates}},
  \bibinfo {author} {\bibfnamefont {G.}~\bibnamefont {Pizzi}}, \bibinfo
  {author} {\bibfnamefont {Y.-S.}\ \bibnamefont {Lee}}, \bibinfo {author}
  {\bibfnamefont {I.}~\bibnamefont {Souza}}, \bibinfo {author} {\bibfnamefont
  {D.}~\bibnamefont {Vanderbilt}},\ and\ \bibinfo {author} {\bibfnamefont
  {N.}~\bibnamefont {Marzari}},\ }\href@noop {} {\bibfield  {journal} {\bibinfo
   {journal} {Comput. Phys. Commun.}\ }\textbf {\bibinfo {volume} {185}},\
  \bibinfo {pages} {2309} (\bibinfo {year} {2014})}\BibitemShut {NoStop}%
\bibitem [{\citenamefont {Korotin}\ \emph {et~al.}(2015)\citenamefont
  {Korotin}, \citenamefont {Mazurenko}, \citenamefont {Anisimov},\ and\
  \citenamefont {Streltsov}}]{korotin2015calculation}%
  \BibitemOpen
  \bibfield  {author} {\bibinfo {author} {\bibfnamefont {D.~M.}\ \bibnamefont
  {Korotin}}, \bibinfo {author} {\bibfnamefont {V.}~\bibnamefont {Mazurenko}},
  \bibinfo {author} {\bibfnamefont {V.}~\bibnamefont {Anisimov}},\ and\
  \bibinfo {author} {\bibfnamefont {S.}~\bibnamefont {Streltsov}},\ }\href@noop
  {} {\bibfield  {journal} {\bibinfo  {journal} {Phys. Rev. B}\ }\textbf
  {\bibinfo {volume} {91}},\ \bibinfo {pages} {224405} (\bibinfo {year}
  {2015})}\BibitemShut {NoStop}%
\bibitem [{\citenamefont {Zhang}\ \emph {et~al.}(2020)\citenamefont {Zhang},
  \citenamefont {Yang},\ and\ \citenamefont {Zhang}}]{zhang2020self}%
  \BibitemOpen
  \bibfield  {author} {\bibinfo {author} {\bibfnamefont {G.-M.}\ \bibnamefont
  {Zhang}}, \bibinfo {author} {\bibfnamefont {Y.-f.}\ \bibnamefont {Yang}},\
  and\ \bibinfo {author} {\bibfnamefont {F.-C.}\ \bibnamefont {Zhang}},\
  }\href@noop {} {\bibfield  {journal} {\bibinfo  {journal} {Phys. Rev. B}\
  }\textbf {\bibinfo {volume} {101}},\ \bibinfo {pages} {020501} (\bibinfo
  {year} {2020})}\BibitemShut {NoStop}%
\bibitem [{\citenamefont {Sandratskii}(1998)}]{sandratskii1998noncollinear}%
  \BibitemOpen
  \bibfield  {author} {\bibinfo {author} {\bibfnamefont {L.}~\bibnamefont
  {Sandratskii}},\ }\href@noop {} {\bibfield  {journal} {\bibinfo  {journal}
  {Adv. Phys.}\ }\textbf {\bibinfo {volume} {47}},\ \bibinfo {pages} {91}
  (\bibinfo {year} {1998})}\BibitemShut {NoStop}%
\bibitem [{\citenamefont {Liechtenstein}\ \emph {et~al.}(1987)\citenamefont
  {Liechtenstein}, \citenamefont {Katsnelson}, \citenamefont {Antropov},\ and\
  \citenamefont {Gubanov}}]{liechtenstein1987local}%
  \BibitemOpen
  \bibfield  {author} {\bibinfo {author} {\bibfnamefont {A.~I.}\ \bibnamefont
  {Liechtenstein}}, \bibinfo {author} {\bibfnamefont {M.}~\bibnamefont
  {Katsnelson}}, \bibinfo {author} {\bibfnamefont {V.}~\bibnamefont
  {Antropov}},\ and\ \bibinfo {author} {\bibfnamefont {V.}~\bibnamefont
  {Gubanov}},\ }\href@noop {} {\bibfield  {journal} {\bibinfo  {journal} {J.
  Magn. Magn. Mater.}\ }\textbf {\bibinfo {volume} {67}},\ \bibinfo {pages}
  {65} (\bibinfo {year} {1987})}\BibitemShut {NoStop}%
\bibitem [{\citenamefont {Si}\ \emph {et~al.}(1993)\citenamefont {Si},
  \citenamefont {Zha}, \citenamefont {Levin},\ and\ \citenamefont
  {Lu}}]{si1993comparison}%
  \BibitemOpen
  \bibfield  {author} {\bibinfo {author} {\bibfnamefont {Q.}~\bibnamefont
  {Si}}, \bibinfo {author} {\bibfnamefont {Y.}~\bibnamefont {Zha}}, \bibinfo
  {author} {\bibfnamefont {K.}~\bibnamefont {Levin}},\ and\ \bibinfo {author}
  {\bibfnamefont {J.}~\bibnamefont {Lu}},\ }\href@noop {} {\bibfield  {journal}
  {\bibinfo  {journal} {Phys. Rev. B}\ }\textbf {\bibinfo {volume} {47}},\
  \bibinfo {pages} {9055} (\bibinfo {year} {1993})}\BibitemShut {NoStop}%
\bibitem [{\citenamefont {Shraiman}\ and\ \citenamefont
  {Siggia}(1989)}]{shraiman1989spiral}%
  \BibitemOpen
  \bibfield  {author} {\bibinfo {author} {\bibfnamefont {B.~I.}\ \bibnamefont
  {Shraiman}}\ and\ \bibinfo {author} {\bibfnamefont {E.~D.}\ \bibnamefont
  {Siggia}},\ }\href@noop {} {\bibfield  {journal} {\bibinfo  {journal} {Phys.
  Rev. Lett.}\ }\textbf {\bibinfo {volume} {62}},\ \bibinfo {pages} {1564}
  (\bibinfo {year} {1989})}\BibitemShut {NoStop}%
\bibitem [{\citenamefont {Sushkov}\ and\ \citenamefont
  {Kotov}(2004)}]{sushkov2004superconducting}%
  \BibitemOpen
  \bibfield  {author} {\bibinfo {author} {\bibfnamefont {O.~P.}\ \bibnamefont
  {Sushkov}}\ and\ \bibinfo {author} {\bibfnamefont {V.~N.}\ \bibnamefont
  {Kotov}},\ }\href@noop {} {\bibfield  {journal} {\bibinfo  {journal} {Phys.
  Rev. B}\ }\textbf {\bibinfo {volume} {70}},\ \bibinfo {pages} {024503}
  (\bibinfo {year} {2004})}\BibitemShut {NoStop}%
\bibitem [{\citenamefont {Tranquada}\ \emph {et~al.}(1995)\citenamefont
  {Tranquada}, \citenamefont {Sternlieb}, \citenamefont {Axe}, \citenamefont
  {Nakamura},\ and\ \citenamefont {Uchida}}]{tranquada1995evidence}%
  \BibitemOpen
  \bibfield  {author} {\bibinfo {author} {\bibfnamefont {J.}~\bibnamefont
  {Tranquada}}, \bibinfo {author} {\bibfnamefont {B.}~\bibnamefont
  {Sternlieb}}, \bibinfo {author} {\bibfnamefont {J.}~\bibnamefont {Axe}},
  \bibinfo {author} {\bibfnamefont {Y.}~\bibnamefont {Nakamura}},\ and\
  \bibinfo {author} {\bibfnamefont {S.-i.}\ \bibnamefont {Uchida}},\
  }\href@noop {} {\bibfield  {journal} {\bibinfo  {journal} {Nature}\ }\textbf
  {\bibinfo {volume} {375}},\ \bibinfo {pages} {561} (\bibinfo {year}
  {1995})}\BibitemShut {NoStop}%
\bibitem [{\citenamefont {Chubukov}\ and\ \citenamefont
  {Musaelian}(1995)}]{chubukov1995magnetic}%
  \BibitemOpen
  \bibfield  {author} {\bibinfo {author} {\bibfnamefont {A.~V.}\ \bibnamefont
  {Chubukov}}\ and\ \bibinfo {author} {\bibfnamefont {K.~A.}\ \bibnamefont
  {Musaelian}},\ }\href@noop {} {\bibfield  {journal} {\bibinfo  {journal}
  {Phys. Rev. B}\ }\textbf {\bibinfo {volume} {51}},\ \bibinfo {pages} {12605}
  (\bibinfo {year} {1995})}\BibitemShut {NoStop}%
\bibitem [{\citenamefont {Scalapino}(1995)}]{scalapino1995case}%
  \BibitemOpen
  \bibfield  {author} {\bibinfo {author} {\bibfnamefont {D.~J.}\ \bibnamefont
  {Scalapino}},\ }\href@noop {} {\bibfield  {journal} {\bibinfo  {journal}
  {Phys. Rep.}\ }\textbf {\bibinfo {volume} {250}},\ \bibinfo {pages} {329}
  (\bibinfo {year} {1995})}\BibitemShut {NoStop}%
\bibitem [{\citenamefont {Emery}\ and\ \citenamefont
  {Kivelson}(1993)}]{emery1993frustrated}%
  \BibitemOpen
  \bibfield  {author} {\bibinfo {author} {\bibfnamefont {V.~J.}\ \bibnamefont
  {Emery}}\ and\ \bibinfo {author} {\bibfnamefont {S.}~\bibnamefont
  {Kivelson}},\ }\href@noop {} {\bibfield  {journal} {\bibinfo  {journal}
  {Physica C Supercond.}\ }\textbf {\bibinfo {volume} {209}},\ \bibinfo {pages}
  {597} (\bibinfo {year} {1993})}\BibitemShut {NoStop}%
\bibitem [{\citenamefont {Vojta}(2009)}]{vojta2009lattice}%
  \BibitemOpen
  \bibfield  {author} {\bibinfo {author} {\bibfnamefont {M.}~\bibnamefont
  {Vojta}},\ }\href@noop {} {\bibfield  {journal} {\bibinfo  {journal} {Adv.
  Phys.}\ }\textbf {\bibinfo {volume} {58}},\ \bibinfo {pages} {699} (\bibinfo
  {year} {2009})}\BibitemShut {NoStop}%
\end{thebibliography}%
\end{document}